\documentclass[twocolumn,superscriptaddress,prd,showkeys,showpacs,nofootinbib]{revtex4-1}

\usepackage{graphicx}
\usepackage[colorlinks = true,
            linkcolor = cyan,
            urlcolor  = blue,
            citecolor = red,
            anchorcolor = blue]{hyperref}\usepackage{color}
\usepackage{amssymb}
\usepackage[nointegrals]{wasysym}
\usepackage{amsthm}
\usepackage{textcomp}
\usepackage{mathtools}
\usepackage{lineno}

\usepackage{comment}
\usepackage[separate-uncertainty,retain-explicit-plus,per-mode = symbol]{siunitx}
\usepackage{url}

\newcommand{\kr}{\mbox{$^{83m}$Kr}}

\newcommand{\us}{$\mu$s}
\newcommand{\el}{e$^{-}$}

\begin{document}

\title{Investigation of background electron emission in the LUX detector}


\author{D.S.~Akerib} \affiliation{SLAC National Accelerator Laboratory, 2575 Sand Hill Road, Menlo Park, CA 94205, USA} \affiliation{Kavli Institute for Particle Astrophysics and Cosmology, Stanford University, 452 Lomita Mall, Stanford, CA 94309, USA} 
\author{S.~Alsum} \affiliation{University of Wisconsin-Madison, Department of Physics, 1150 University Ave., Madison, WI 53706, USA}  
\author{H.M.~Ara\'{u}jo} \affiliation{Imperial College London, High Energy Physics, Blackett Laboratory, London SW7 2BZ, United Kingdom}  
\author{X.~Bai} \affiliation{South Dakota School of Mines and Technology, 501 East St Joseph St., Rapid City, SD 57701, USA}  
\author{J.~Balajthy} \affiliation{University of California Davis, Department of Physics, One Shields Ave., Davis, CA 95616, USA}  
\author{A.~Baxter} \affiliation{University of Liverpool, Department of Physics, Liverpool L69 7ZE, UK}  
\author{E.P.~Bernard} \affiliation{University of California Berkeley, Department of Physics, Berkeley, CA 94720, USA}  
\author{A.~Bernstein} \affiliation{Lawrence Livermore National Laboratory, 7000 East Ave., Livermore, CA 94551, USA}  
\author{T.P.~Biesiadzinski} \affiliation{SLAC National Accelerator Laboratory, 2575 Sand Hill Road, Menlo Park, CA 94205, USA} \affiliation{Kavli Institute for Particle Astrophysics and Cosmology, Stanford University, 452 Lomita Mall, Stanford, CA 94309, USA} 
\author{E.M.~Boulton} \affiliation{University of California Berkeley, Department of Physics, Berkeley, CA 94720, USA} \affiliation{Lawrence Berkeley National Laboratory, 1 Cyclotron Rd., Berkeley, CA 94720, USA} \affiliation{Yale University, Department of Physics, 217 Prospect St., New Haven, CT 06511, USA}
\author{B.~Boxer} \affiliation{University of Liverpool, Department of Physics, Liverpool L69 7ZE, UK}  
\author{P.~Br\'as} \affiliation{LIP-Coimbra, Department of Physics, University of Coimbra, Rua Larga, 3004-516 Coimbra, Portugal}  
\author{S.~Burdin} \affiliation{University of Liverpool, Department of Physics, Liverpool L69 7ZE, UK}  
\author{D.~Byram} \affiliation{University of South Dakota, Department of Physics, 414E Clark St., Vermillion, SD 57069, USA} \affiliation{South Dakota Science and Technology Authority, Sanford Underground Research Facility, Lead, SD 57754, USA} 
\author{M.C.~Carmona-Benitez} \affiliation{Pennsylvania State University, Department of Physics, 104 Davey Lab, University Park, PA  16802-6300, USA}  
\author{C.~Chan} \affiliation{Brown University, Department of Physics, 182 Hope St., Providence, RI 02912, USA}  
\author{J.E.~Cutter} \affiliation{University of California Davis, Department of Physics, One Shields Ave., Davis, CA 95616, USA}  
\author{L.~de\,Viveiros}  \affiliation{Pennsylvania State University, Department of Physics, 104 Davey Lab, University Park, PA  16802-6300, USA}  
\author{E.~Druszkiewicz} \affiliation{University of Rochester, Department of Physics and Astronomy, Rochester, NY 14627, USA}  
\author{A.~Fan} \affiliation{SLAC National Accelerator Laboratory, 2575 Sand Hill Road, Menlo Park, CA 94205, USA} \affiliation{Kavli Institute for Particle Astrophysics and Cosmology, Stanford University, 452 Lomita Mall, Stanford, CA 94309, USA} 
\author{S.~Fiorucci} \affiliation{Lawrence Berkeley National Laboratory, 1 Cyclotron Rd., Berkeley, CA 94720, USA} \affiliation{Brown University, Department of Physics, 182 Hope St., Providence, RI 02912, USA} 
\author{R.J.~Gaitskell} \affiliation{Brown University, Department of Physics, 182 Hope St., Providence, RI 02912, USA}  
\author{C.~Ghag} \affiliation{Department of Physics and Astronomy, University College London, Gower Street, London WC1E 6BT, United Kingdom}  
\author{M.G.D.~Gilchriese} \affiliation{Lawrence Berkeley National Laboratory, 1 Cyclotron Rd., Berkeley, CA 94720, USA}  
\author{C.~Gwilliam} \affiliation{University of Liverpool, Department of Physics, Liverpool L69 7ZE, UK}  
\author{C.R.~Hall} \affiliation{University of Maryland, Department of Physics, College Park, MD 20742, USA}  
\author{S.J.~Haselschwardt} \affiliation{University of California Santa Barbara, Department of Physics, Santa Barbara, CA 93106, USA}  
\author{S.A.~Hertel} \affiliation{University of Massachusetts, Amherst Center for Fundamental Interactions and Department of Physics, Amherst, MA 01003-9337 USA} \affiliation{Lawrence Berkeley National Laboratory, 1 Cyclotron Rd., Berkeley, CA 94720, USA} 
\author{D.P.~Hogan} \affiliation{University of California Berkeley, Department of Physics, Berkeley, CA 94720, USA}  
\author{M.~Horn} \affiliation{South Dakota Science and Technology Authority, Sanford Underground Research Facility, Lead, SD 57754, USA} \affiliation{University of California Berkeley, Department of Physics, Berkeley, CA 94720, USA} 
\author{D.Q.~Huang} \affiliation{Brown University, Department of Physics, 182 Hope St., Providence, RI 02912, USA}  
\author{C.M.~Ignarra} \affiliation{SLAC National Accelerator Laboratory, 2575 Sand Hill Road, Menlo Park, CA 94205, USA} \affiliation{Kavli Institute for Particle Astrophysics and Cosmology, Stanford University, 452 Lomita Mall, Stanford, CA 94309, USA} 
\author{R.G.~Jacobsen} \affiliation{University of California Berkeley, Department of Physics, Berkeley, CA 94720, USA}  
\author{O.~Jahangir} \affiliation{Department of Physics and Astronomy, University College London, Gower Street, London WC1E 6BT, United Kingdom}  
\author{W.~Ji} \affiliation{SLAC National Accelerator Laboratory, 2575 Sand Hill Road, Menlo Park, CA 94205, USA} \affiliation{Kavli Institute for Particle Astrophysics and Cosmology, Stanford University, 452 Lomita Mall, Stanford, CA 94309, USA} 
\author{K.~Kamdin} \affiliation{University of California Berkeley, Department of Physics, Berkeley, CA 94720, USA} \affiliation{Lawrence Berkeley National Laboratory, 1 Cyclotron Rd., Berkeley, CA 94720, USA} 
\author{K.~Kazkaz} \affiliation{Lawrence Livermore National Laboratory, 7000 East Ave., Livermore, CA 94551, USA}  
\author{D.~Khaitan} \affiliation{University of Rochester, Department of Physics and Astronomy, Rochester, NY 14627, USA}  
\author{E.V.~Korolkova} \affiliation{University of Sheffield, Department of Physics and Astronomy, Sheffield, S3 7RH, United Kingdom}  
\author{S.~Kravitz} \affiliation{Lawrence Berkeley National Laboratory, 1 Cyclotron Rd., Berkeley, CA 94720, USA}  
\author{V.A.~Kudryavtsev} \affiliation{University of Sheffield, Department of Physics and Astronomy, Sheffield, S3 7RH, United Kingdom}  
\author{E.~Leason} \affiliation{SUPA, School of Physics and Astronomy, University of Edinburgh, Edinburgh EH9 3FD, United Kingdom}  
\author{B.G.~Lenardo} \affiliation{University of California Davis, Department of Physics, One Shields Ave., Davis, CA 95616, USA} \affiliation{Lawrence Livermore National Laboratory, 7000 East Ave., Livermore, CA 94551, USA} 
\author{K.T.~Lesko} \affiliation{Lawrence Berkeley National Laboratory, 1 Cyclotron Rd., Berkeley, CA 94720, USA}  
\author{J.~Liao} \affiliation{Brown University, Department of Physics, 182 Hope St., Providence, RI 02912, USA}  
\author{J.~Lin} \affiliation{University of California Berkeley, Department of Physics, Berkeley, CA 94720, USA}  
\author{A.~Lindote} \affiliation{LIP-Coimbra, Department of Physics, University of Coimbra, Rua Larga, 3004-516 Coimbra, Portugal}  
\author{M.I.~Lopes} \affiliation{LIP-Coimbra, Department of Physics, University of Coimbra, Rua Larga, 3004-516 Coimbra, Portugal}  
\author{A.~Manalaysay} \affiliation{Lawrence Berkeley National Laboratory, 1 Cyclotron Rd., Berkeley, CA 94720, USA} \affiliation{University of California Davis, Department of Physics, One Shields Ave., Davis, CA 95616, USA} 
\author{R.L.~Mannino} \affiliation{Texas A \& M University, Department of Physics, College Station, TX 77843, USA} \affiliation{University of Wisconsin-Madison, Department of Physics, 1150 University Ave., Madison, WI 53706, USA} 
\author{N.~Marangou} \affiliation{Imperial College London, High Energy Physics, Blackett Laboratory, London SW7 2BZ, United Kingdom}  
\author{D.N.~McKinsey} \affiliation{University of California Berkeley, Department of Physics, Berkeley, CA 94720, USA} \affiliation{Lawrence Berkeley National Laboratory, 1 Cyclotron Rd., Berkeley, CA 94720, USA} 
\author{D.-M.~Mei} \affiliation{University of South Dakota, Department of Physics, 414E Clark St., Vermillion, SD 57069, USA}  
\author{M.~Moongweluwan} \affiliation{University of Rochester, Department of Physics and Astronomy, Rochester, NY 14627, USA}  
\author{J.A.~Morad} \affiliation{University of California Davis, Department of Physics, One Shields Ave., Davis, CA 95616, USA}  
\author{A.St.J.~Murphy} \affiliation{SUPA, School of Physics and Astronomy, University of Edinburgh, Edinburgh EH9 3FD, United Kingdom}  
\author{A.~Naylor} \affiliation{University of Sheffield, Department of Physics and Astronomy, Sheffield, S3 7RH, United Kingdom}  
\author{C.~Nehrkorn} \affiliation{University of California Santa Barbara, Department of Physics, Santa Barbara, CA 93106, USA}  
\author{H.N.~Nelson} \affiliation{University of California Santa Barbara, Department of Physics, Santa Barbara, CA 93106, USA}  
\author{F.~Neves} \affiliation{LIP-Coimbra, Department of Physics, University of Coimbra, Rua Larga, 3004-516 Coimbra, Portugal}  
\author{A.~Nilima} \affiliation{SUPA, School of Physics and Astronomy, University of Edinburgh, Edinburgh EH9 3FD, United Kingdom}  
\author{K.C.~Oliver-Mallory} \affiliation{University of California Berkeley, Department of Physics, Berkeley, CA 94720, USA} \affiliation{Lawrence Berkeley National Laboratory, 1 Cyclotron Rd., Berkeley, CA 94720, USA} 
\author{K.J.~Palladino} \affiliation{University of Wisconsin-Madison, Department of Physics, 1150 University Ave., Madison, WI 53706, USA}  
\author{E.K.~Pease} \affiliation{University of California Berkeley, Department of Physics, Berkeley, CA 94720, USA} \affiliation{Lawrence Berkeley National Laboratory, 1 Cyclotron Rd., Berkeley, CA 94720, USA} 
\author{Q.~Riffard} \affiliation{University of California Berkeley, Department of Physics, Berkeley, CA 94720, USA} \affiliation{Lawrence Berkeley National Laboratory, 1 Cyclotron Rd., Berkeley, CA 94720, USA} 
\author{G.R.C.~Rischbieter} \affiliation{University at Albany, State University of New York, Department of Physics, 1400 Washington Ave., Albany, NY 12222, USA}  
\author{C.~Rhyne} \affiliation{Brown University, Department of Physics, 182 Hope St., Providence, RI 02912, USA}  
\author{P.~Rossiter} \affiliation{University of Sheffield, Department of Physics and Astronomy, Sheffield, S3 7RH, United Kingdom}  
\author{S.~Shaw} \affiliation{University of California Santa Barbara, Department of Physics, Santa Barbara, CA 93106, USA} \affiliation{Department of Physics and Astronomy, University College London, Gower Street, London WC1E 6BT, United Kingdom} 
\author{T.A.~Shutt} \affiliation{SLAC National Accelerator Laboratory, 2575 Sand Hill Road, Menlo Park, CA 94205, USA} \affiliation{Kavli Institute for Particle Astrophysics and Cosmology, Stanford University, 452 Lomita Mall, Stanford, CA 94309, USA} 
\author{C.~Silva} \affiliation{LIP-Coimbra, Department of Physics, University of Coimbra, Rua Larga, 3004-516 Coimbra, Portugal}  
\author{M.~Solmaz} \affiliation{University of California Santa Barbara, Department of Physics, Santa Barbara, CA 93106, USA}  
\author{V.N.~Solovov} \affiliation{LIP-Coimbra, Department of Physics, University of Coimbra, Rua Larga, 3004-516 Coimbra, Portugal}  
\author{P.~Sorensen} \affiliation{Lawrence Berkeley National Laboratory, 1 Cyclotron Rd., Berkeley, CA 94720, USA}  
\author{T.J.~Sumner} \affiliation{Imperial College London, High Energy Physics, Blackett Laboratory, London SW7 2BZ, United Kingdom}  
\author{M.~Szydagis} \affiliation{University at Albany, State University of New York, Department of Physics, 1400 Washington Ave., Albany, NY 12222, USA}  
\author{D.J.~Taylor} \affiliation{South Dakota Science and Technology Authority, Sanford Underground Research Facility, Lead, SD 57754, USA}  
\author{R.~Taylor} \affiliation{Imperial College London, High Energy Physics, Blackett Laboratory, London SW7 2BZ, United Kingdom}  
\author{W.C.~Taylor} \affiliation{Brown University, Department of Physics, 182 Hope St., Providence, RI 02912, USA}  
\author{B.P.~Tennyson} \affiliation{Yale University, Department of Physics, 217 Prospect St., New Haven, CT 06511, USA}  
\author{P.A.~Terman} \affiliation{Texas A \& M University, Department of Physics, College Station, TX 77843, USA}  
\author{D.R.~Tiedt} \affiliation{University of Maryland, Department of Physics, College Park, MD 20742, USA}  
\author{W.H.~To} \affiliation{California State University Stanislaus, Department of Physics, 1 University Circle, Turlock, CA 95382, USA}  
\author{L.~Tvrznikova} \affiliation{University of California Berkeley, Department of Physics, Berkeley, CA 94720, USA} \affiliation{Lawrence Berkeley National Laboratory, 1 Cyclotron Rd., Berkeley, CA 94720, USA} \affiliation{Yale University, Department of Physics, 217 Prospect St., New Haven, CT 06511, USA}
\author{U.~Utku} \affiliation{Department of Physics and Astronomy, University College London, Gower Street, London WC1E 6BT, United Kingdom}  
\author{S.~Uvarov} \affiliation{University of California Davis, Department of Physics, One Shields Ave., Davis, CA 95616, USA}  
\author{A.~Vacheret} \affiliation{Imperial College London, High Energy Physics, Blackett Laboratory, London SW7 2BZ, United Kingdom}  
\author{V.~Velan} \affiliation{University of California Berkeley, Department of Physics, Berkeley, CA 94720, USA}  
\author{R.C.~Webb} \affiliation{Texas A \& M University, Department of Physics, College Station, TX 77843, USA}  
\author{J.T.~White} \affiliation{Texas A \& M University, Department of Physics, College Station, TX 77843, USA}  
\author{T.J.~Whitis} \affiliation{SLAC National Accelerator Laboratory, 2575 Sand Hill Road, Menlo Park, CA 94205, USA} \affiliation{Kavli Institute for Particle Astrophysics and Cosmology, Stanford University, 452 Lomita Mall, Stanford, CA 94309, USA} 
\author{M.S.~Witherell} \affiliation{Lawrence Berkeley National Laboratory, 1 Cyclotron Rd., Berkeley, CA 94720, USA}  
\author{F.L.H.~Wolfs} \affiliation{University of Rochester, Department of Physics and Astronomy, Rochester, NY 14627, USA}  
\author{D.~Woodward} \affiliation{Pennsylvania State University, Department of Physics, 104 Davey Lab, University Park, PA  16802-6300, USA}  
\author{J.~Xu} \email[Corresponding author, ] {xu12@llnl.gov} \affiliation{Lawrence Livermore National Laboratory, 7000 East Ave., Livermore, CA 94551, USA}  
\author{C.~Zhang} \affiliation{University of South Dakota, Department of Physics, 414E Clark St., Vermillion, SD 57069, USA}  



\begin{abstract}
\noindent Dual-phase xenon detectors, as currently used in direct detection dark matter experiments, have observed elevated rates of background electron events in the low energy region. 
While this background negatively impacts detector performance in various ways, its origins have only been partially studied. 
In this paper we report a systematic investigation of the electron pathologies observed in the LUX dark matter experiment. 
We characterize different electron populations based on their emission intensities and their correlations with preceding energy depositions in the detector.
By studying the background under different experimental conditions, 
we identified the leading emission mechanisms, including photoionization and the photoelectric effect induced by the xenon luminescence, delayed emission of electrons trapped under the liquid surface, capture and release of drifting electrons by impurities, and grid electron emission. 
We discuss how these backgrounds can be mitigated in LUX and future xenon-based dark matter experiments. 

\end{abstract}


\maketitle




\section{Introduction}
\label{sec:intro}

The dual-phase xenon time projection chamber (TPC) is one of the few particle detection technologies to have demonstrated sensitivities to single ionization electrons~\cite{ZEPLIN2008_SE, Sangiorgio2013_Ar37, SENSEI2017_SE, CDMS2018_HVSE}. 
In liquid xenon, it only takes $\sim$15~eV of energy for electron recoils~\cite{Boulton2017_Ar37, LUX2017_Xe127}, or $\sim$250~eV for nuclear recoils~\cite{Xu2019_NR}, to produce one ionization electron. 
Through proportional electroluminescence (EL) amplification in xenon gas, an electron can produce hundreds to thousands of secondary photons~\cite{Monteiro2007_XeScint}. 
A typical xenon TPC used in dark matter search experiments can detect a few dozen EL photons for each electron~\cite{ZEPLIN2011_SE, XENON2011_S2Only, XENON2014_SE, LUX2018_Run3PRD}, 
and higher electron gain values of $\gtrsim$100 photoelectrons (PHEs) per electron have also been demonstrated~\cite{PIXeY2018_EEE, Xu2019_EEE}. 
The observation of single electron (SE) events not only provides an {\it in-situ} calibration for these experiments, 
but has also enabled them to attain world-leading sensitivities to GeV- and sub-GeV-mass dark matter candidates, substantially below the mass range targeted by these detectors~\cite{XENON2011_S2Only, Essig2012_SubGeVXENON10, Essig2017_SubGeVXENON100}. 

Despite the exceptionally low background rates achieved in these underground experiments for energy depositions at the keV level and above, xenon TPCs exhibit elevated rates of 
electron backgrounds similar to those expected from small energy depositions~\cite{XENON2011_S2Only, Essig2012_SubGeVXENON10, Sorensen2017_EEE, SorensenElectronBG_2018}. 
This electron background negatively impacts the performance of xenon TPCs. 
For example, spurious electron (or few-electron) pulses can be incorrectly identified as true ionization events, or part of such events, causing inaccurate energy estimation and compromising detector energy resolution~\cite{LUX2017_Yield}. 
In addition, due to their high rates and large pulse areas, these electrons generate excessive triggers and pose a significant burden on the data acquisition, storage and processing systems of xenon TPC experiments. 
Most importantly, this background impairs the ability of xenon TPCs to search for ultra-low energy interactions to which these detectors are otherwise sensitive. 
This problem is most notable in rare event searches that rely on the high-gain ionization signals when scintillation signals are absent or at the detection limit~\cite{XENON2011_S2Only, Essig2012_SubGeVXENON10,XENON1T2019_S2Only, LUX2020_DPE}. 
Although preliminary successes have been demonstrated, 
the excess rate of ionization-like background has so far limited further improvement of the low-energy sensitivity of Xe TPCs.

Several authors have studied electron emission in xenon TPCs, and developed viable hypotheses that explain certain background populations~\cite{Edwards2009_Thesis, ZEPLIN2011_SE, XENON2014_SE, Chapman_PhDThesis, Sorensen2017_EEE, SorensenElectronBG_2018, LZ2018_WireTest, Akimov2019_REDEBg}. 
In this work, we strive to obtain a coherent picture of these background effects through a systematic investigation of all observed electron pathologies in the Large Underground Xenon (LUX) dark matter experiment~\cite{LUX2012_Detector}. 
The LUX xenon TPC was operated 4850 feet below the surface at the Sanford Underground Research Facility (SURF) between 2013 and 2016, 
and demonstrated one of the lowest background rates in dark matter direct search experiments~\cite{LUX2016_Run3_4}. 

The LUX experiment produced a wealth of information needed for a thorough electron background study. 
First, LUX achieved a low background rate of a few counts per second (CPS) in the 250~kg active xenon mass from internal and external radioactivity, 
which leaves large time intervals between particle interactions for pathological electron emission to be studied. 
Second, the LUX data acquisition system allowed all PMT outputs, including PHE and SE pulses, to be continuously recorded for investigation of low energy events~\cite{LUX2012_DAQ}. 
Third, over the three-year life span, LUX underwent a range of operating conditions, including various source calibrations, evolving impurity concentrations in the liquid xenon, and distinct electric field configurations throughout the active volume~\cite{LUX2017_3DField}. 
These expansive data sets enable correlations between operation conditions and electron background behaviors to be studied, so that different hypotheses of electron emission mechanism can be tested. 

This paper is organized as follows: 
Section~\ref{sec:source} reviews the possible charge production and migration mechanisms in dual-phase xenon TPCs; 
Section~\ref{sec:overview} to \ref{sec:me} describe each population of the electron background observed in LUX, characterize their emission behaviors under varying experimental conditions, and discuss possible mechanisms that may be responsible for the production of these electrons.  
Section~\ref{sec:concl} summarizes the major findings of this work and discusses the implications to dark matter searches. 

\section{Ionization phenomena in LUX}
\label{sec:source}

\begin{figure}[h!]
\centering
\includegraphics[width=.45\textwidth]{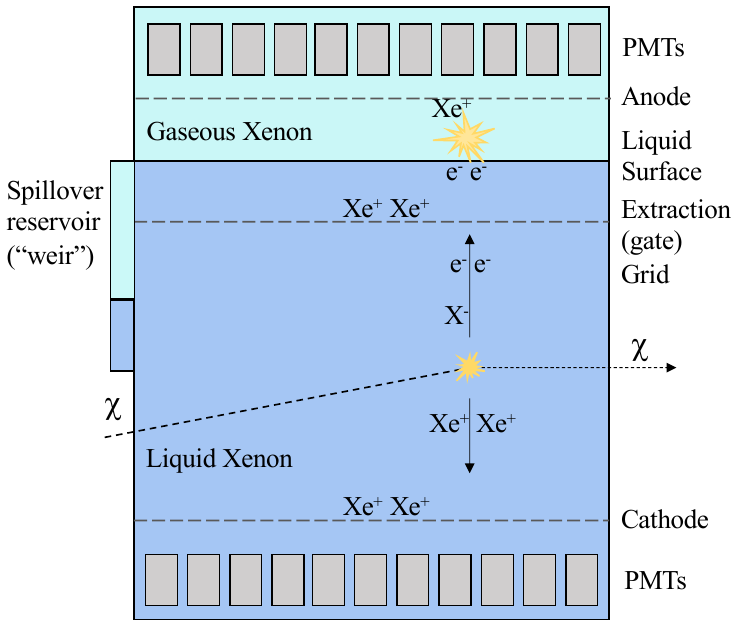}
\caption{A schematic illustration of a dual-phase xenon TPC such as LUX, including the major charge production and migration schemes inside the detector.
In LUX, all liquid xenon flows into a weir before going into the purification system. }
\label{fig:tpc}
\end{figure}

The LUX detector contained 370~kg of pure xenon in a double-walled cryostat, which was hosted in a 7.6~m (diameter) by 6.1~m (height) water tank in the Davis Cavern of SURF. 
The central 250~kg of the liquid xenon, enclosed in an electric field cage, defined the active target volume of the TPC. 
Particle interactions with liquid xenon produce both scintillation photons and ionization electrons. 
In a LUX-style dual-phase TPC as illustrated in Figure~\ref{fig:tpc}, scintillation photons (the so-called ``S1'' or primary scintillation signal) are directly collected by two arrays of photomultipliers (PMTs) above and below the active volume. 
Ionization electrons are drifted upward and are extracted into the gas by the applied electric fields in the TPC. 
As electrons accelerate in the gas under the field, they produce secondary EL photons that are collected by the PMTs (referred to as the ``S2'' or secondary EL signal). 
Combining the associated S1 and S2 signals, one can obtain information about the energy, position and interaction type of the events. 
More information about the LUX detector can be found in \cite{LUX2012_Detector}. 

In an ideal scenario, all ionization electrons produced in the liquid xenon would propagate to the gas phase and be detected. 
In reality, however, a fraction of the electrons is lost or temporarily trapped. 
First, electronegative impurities in liquid xenon can capture electrons and cause the detectable ionization signal to decrease exponentially with the drift time of the ionization clouds. 
In the LUX WIMP-search data acquired in 2013 (WS2013), the electron lifetime value was measured to vary between a few hundred microseconds and approximately one millisecond. 
At a typical electron lifetime of 750~\us, approximately 35\% of ionization electrons from interactions near the bottom of the TPC (maximum drift time of 325~\us) were captured by impurities. 
The newly formed negative ions (denoted as ``X$^-$'' in Figure~\ref{fig:tpc}) are expected to drift in the detector under the effect of electric fields and the liquid flow. 
Contrary to electrons that can drift swiftly in liquid xenon (1.5~mm/\us\ at the LUX drift field of 180~V/cm), negative ions have mobilities that are approximately a million times smaller. 
The drift velocity of O$^{-}_{2}$ is calculated to be around 2~mm/s in LUX~\cite{Hilt1994_IonMobilityLXe}, which is even slower than typical liquid convection flows observed in xenon TPCs ($\mathcal{O}$(cm/s))~\cite{Malling2014_Thesis, XENON2017_Rn220}.
Secondly, electrons arriving at the liquid surface can only be extracted into the gas if they have kinetic energy above a certain threshold~\cite{Gushchin1979_EEE, Reininger1982_EinXe, Bolozdynya1995_XeTPC}. 
In WS2013, the electron extraction efficiency was measured to be 49$\pm$3\%, and it was improved to 73$\pm$4\% for LUX data taken from 2014 to 2016 (WS2014--16)~\cite{LUX2016_Run3_4}. 
Unextracted electrons are thus trapped under the liquid surface; 
they may migrate to the wall, or spill into a liquid xenon reservoir (called a ``weir'' in LUX) and are removed from the active detector volume, 
or become captured by impurities and then drift in the liquid. 
Given sufficient excitation energy, both impurity-captured electrons and surface-trapped electrons may be liberated and become a background~\cite{SorensenElectronBG_2018}. 

While charge production and migration in xenon are usually discussed in the context of ionization electrons, 
it is important to also consider the corresponding positive ions produced. 
These positive ion clouds are expected to drift down to the cathode under electric fields, but they will also migrate with the liquid flow, in much the same way as negative ions. 
However, positive xenon ions, or ``holes'', have a higher mobility in liquid xenon than that of negative ions~\cite{Hilt1994_IonMobilityLXe} and are estimated to drift at a velocity of 8~mm/s at the LUX field.
In addition, extracted electrons may produce additional ionization in the gas as they accelerate, especially in the high electric field regions near the anode grid wires;  
these resulting ions then drift down toward the liquid under the effect of applied electric fields, with an estimated drift velocity of 15--20~mm/ms in LUX~\cite{Neves2007_XeGasMobility}. 
If positive ions manage to reach the electrode grids, they can neutralize with the metal, 
or they may accumulate on the surfaces if neutralization is prevented by oxide layers or monolayers of solid xenon on metal surfaces~\cite{Bruch2007_Physisorption}.
Ion neutralization on metal surfaces can produce Auger neutralization electrons to discharge the combination energy~\cite{Monreal2014_Auger}. 

Particle interactions with xenon near detector surfaces lead to very different electric charge behaviors. 
Dielectric materials such as polytetrafluoroethylene (PTFE) reflectors may attract free charges near their surfaces and cause incomplete electron collection. 
Radioactive decays on electrode grid surfaces, where the electric field can reach very high values due to the relatively small surface area, usually result in highly suppressed and obscured scintillation signals and enhanced ionization signals that may be detected as pure charge events. 

In addition to charge production by ionizing particles, free charges may also be generated in LUX from instrumental effects. 
For example, the ultraviolet xenon scintillation and EL photons carry an energy of $\sim$7~eV, which is above the work function of many metals and other species; 
therefore, the photoelectric effect can liberate electrons from the electrode surfaces and certain impurities dissolved in the liquid xenon~\cite{ZEPLIN2011_SE, XENON2014_SE, Huang_PhDThesis}. 
In addition, electrons may be emitted from the cathodic electrode wire surfaces, where the electric field can reach very high values if physical or chemical defects are present~\cite{Bailey2016_Thesis, LZ2018_WireTest}. 
Electron emission from electrodes can lead to high voltage instabilities or even breakdowns and has prevented several TPC experiments from operating at the designed field configurations~\cite{Rebel2014_NLHV, LUX2017_3DField}.

\newpage
\onecolumngrid

\begin{figure}[t!]
\centering
\includegraphics[width=.95\textwidth]{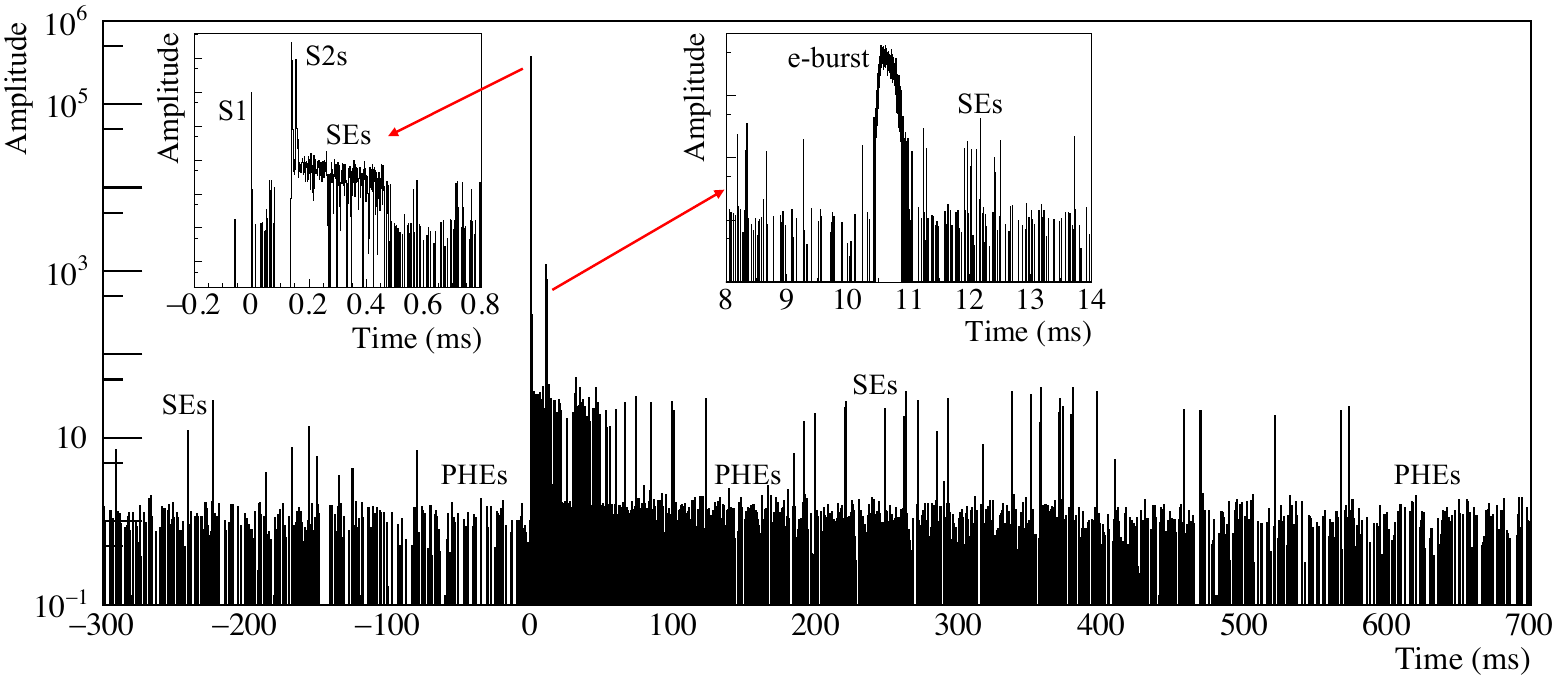}
\caption{A continuous LUX waveform over one second. Within this time window, only one 2.3~MeV gamma ray interacts with LUX but thousands of background electrons are observed to follow the S1 and S2 pulses. 
The insets show zoomed-in views of the S1-S2 event window and a tail window that contains an electron cluster (``e-burst'') and sparse single electron (SE) pulses and photoelectron (PHE) pulses. 
The waveform amplitude corresponds to the integral of calibrated PMT traces in time windows of 2~\us; pulses with amplitudes near 1 are PHEs and pulses with amplitudes around 20 are SEs. 
Due to the coarse binning used in the time axes, typical S1/S2/SE/PHE pulses are seen as individual lines. 
}
\label{fig:wfm}
\end{figure}

\twocolumngrid

\section{Overview of electron background in LUX}
\label{sec:overview}

With a rate of 3--4 CPS in the whole active liquid xenon volume above a few keV, consecutive particle interactions in LUX are typically separated by hundreds of milliseconds. 
These long interaction-free windows allow delayed electron emission to be studied at timescales far exceeding the typical analysis event window (1~ms). 
Figure~\ref{fig:wfm} shows a continuous LUX waveform over a one-second period. 
The interactions of a 2.3~MeV gamma ray with liquid xenon lead to the detection of 9,300 prompt photons and 41,000 electrons. 
Following the S1 and S2 pulses, an increased population of electrons and photons emerge, which lasts for hundreds of milliseconds and into the next interaction event. 
The background comprises mostly sparse SE and PHE pulses, but also contains intense clustered electron emission. 
This apparent time correlation leads us to conclude that these electrons are produced by the prior, relatively high energy interactions that precede them. 
In the following sections, we examine possible production mechanisms for these induced electrons and electron clusters.  

Based on the emission characteristics of the background electrons and their time correlation with preceding events, 
we place them in four categories: 1) photoionization electrons that are detected within hundreds of microseconds after the S1 and S2\unskip~\footnote{Throughout this paper, S2s only refer to EL pulses produced by energy depositions in liquid xenon from radioactivity, excluding pathological electrons.} pulses,
2) clustered electron emission that occurs within tens of milliseconds after S2s, 
3) delayed emission of individual electrons at the millisecond-to-second scale, and 4) electron emission that appears independent of prior interactions. 
In the following sections we will quantitatively describe the electron populations and study their correlations with experimental operation conditions so that connections between these emission processes and the ionization phenomena presented in Section~\ref{sec:source} can be made. 

This work primarily uses LUX WS2013 data unless specified otherwise. 
Unlike previous LUX studies that used the same data set, the analysis framework is redesigned for the efficient identification and parametrization of small pulses such as SEs and PHEs. 
In particular, this work is independent of the LUX event-building system, and treats all recorded PMT waveforms as a nearly continuous stream of pulses.~\unskip\footnote{Dead-time in recorded LUX data is $\mathcal{O}$(0.1\%), or 2-3~ms for every $\sim$2~s data acquisition window.}
Special care was given to baseline corrections, pulse finding and splitting of closely overlapped pulses. 
Thousands of waveforms were visually scanned, based on which we estimate an efficiency of $>$95\% for identifying SE and PHE pulses in the analysis, 
and the efficiency loss is mostly due to misidentification in the high pulse rate regions shortly after large S2s.  
The additional loss of efficiency due to the internal digitizer thresholds is estimated to be 5\% for PHEs~\cite{LUX2012_DAQ}. 
For position reconstruction of S2s and SEs, instead of using the sophisticated Mercury algorithm~\cite{LUX2018_Position}, a less computation-intensive method was developed that uses only the group of 7 neighboring PMTs with the most detected light in the top PMT array (3 or 4 PMTs for events at the perimeter of LUX). 
A comparison to standard LUX results indicates a modest degradation of the position resolution ($\sigma$) from 2.1~cm to 2.9~cm for SE pulses. 
To mitigate PMT saturation effects in the evaluation of S2 sizes, we exclusively use the pulse area observed in the bottom PMT array multiplied by a measured scaling factor to account for the top-array contribution. 
No other corrections are applied to the S2 or S1 areas used in this work unless specified otherwise. 
In all following analyses, we will quantify background descriptions whenever possible, but a full study of uncertainties is beyond the scope of this work. 

\section{Photoionization electrons}
\label{sec:pi}

\begin{figure}[b!]
\centering
\includegraphics[width=.42\textwidth]{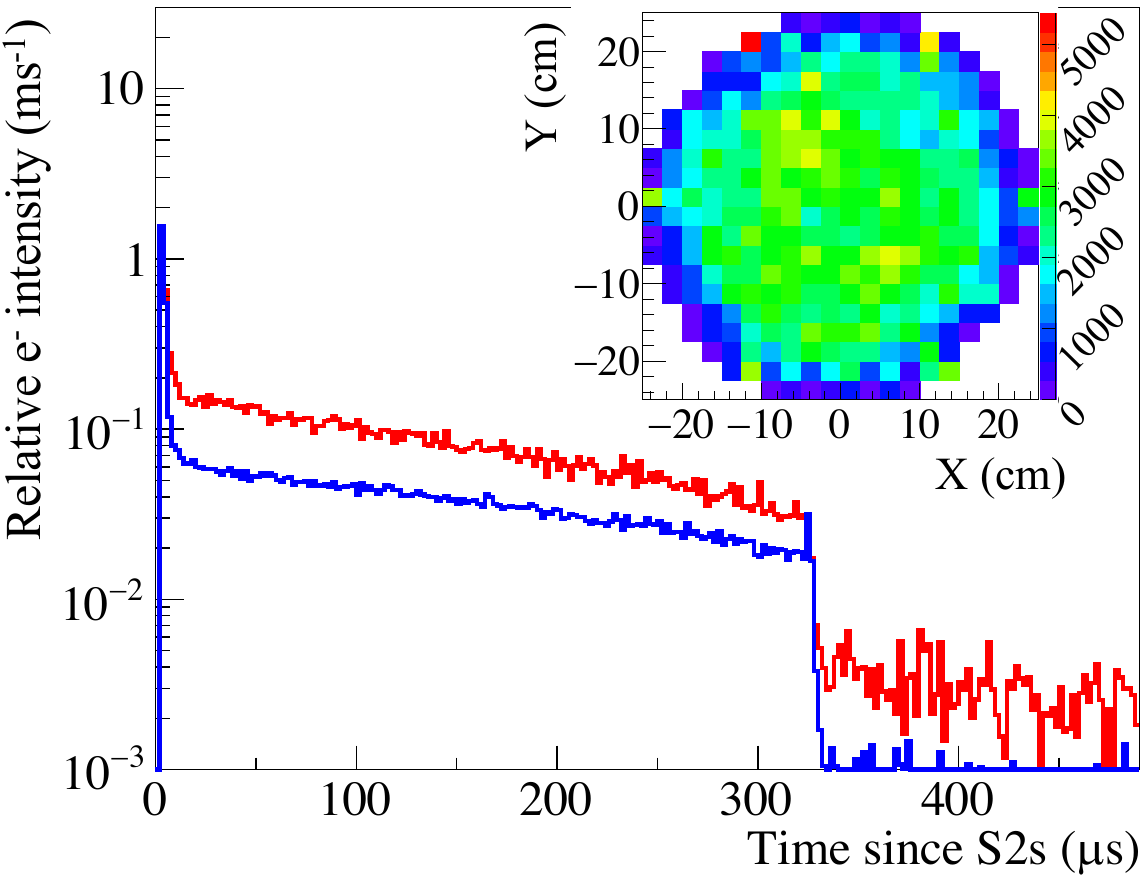}
\caption{The relative intensity of SEs detected after \kr-decay S2s at low (red) and high (blue) electron lifetime values; 
the intensity is defined as the electron pulse area divided by the raw S2 area. 
The peaks near 0 and 325~\us\ are mostly from photoelectric effects on the electrode surfaces, and the continuous distribution in between is from photoionization in the liquid. 
The inset shows the horizontal (X-Y) position distribution of SEs detected within 10--320~\us\ after \kr\ S2s of $r<$10~cm. }
\label{fig:pi_ratexy}
\end{figure}

As illustrated in the inset of Figure~\ref{fig:wfm} (top left), a prominent electron background population is observed immediately after S2 pulses. 
This background is comprised of individual electron pulses with a rate that decreases with time. 
In the case of large S2s, a sharp drop in the electron rate is observed at 325~\us---the maximum drift time in LUX---after the S2 time. 
Figure~\ref{fig:pi_ratexy} shows the arrival time of electron pulses  and their X-Y positions (inset) following \kr\ calibration decay events~\cite{LUX2017_Kr83m}. 
The intensity of electron emission increases linearly with S2 pulse area. 
It also increases with impurity concentration in the liquid xenon, measured by the electron lifetime (inverse of the probability for an electron to be captured by impurities in liquid xenon per unit drift time). 
Based on these observations, the immediate electrons (time delay $<$325~\us) are attributed to the photoionization by S2 light on impurities dissolved in liquid xenon. 
This phenomenon has been observed and discussed in other xenon TPC experiments including ZEPLIN-II~\cite{ZEPLIN2008_SE}, ZEPLIN-III~\cite{ZEPLIN2011_SE} and XENON100~\cite{XENON2014_SE}.

\subsection{Photoionization yield}
\label{sec:pi_yield}

\begin{figure}[h!]
\centering
\includegraphics[width=.49\textwidth]{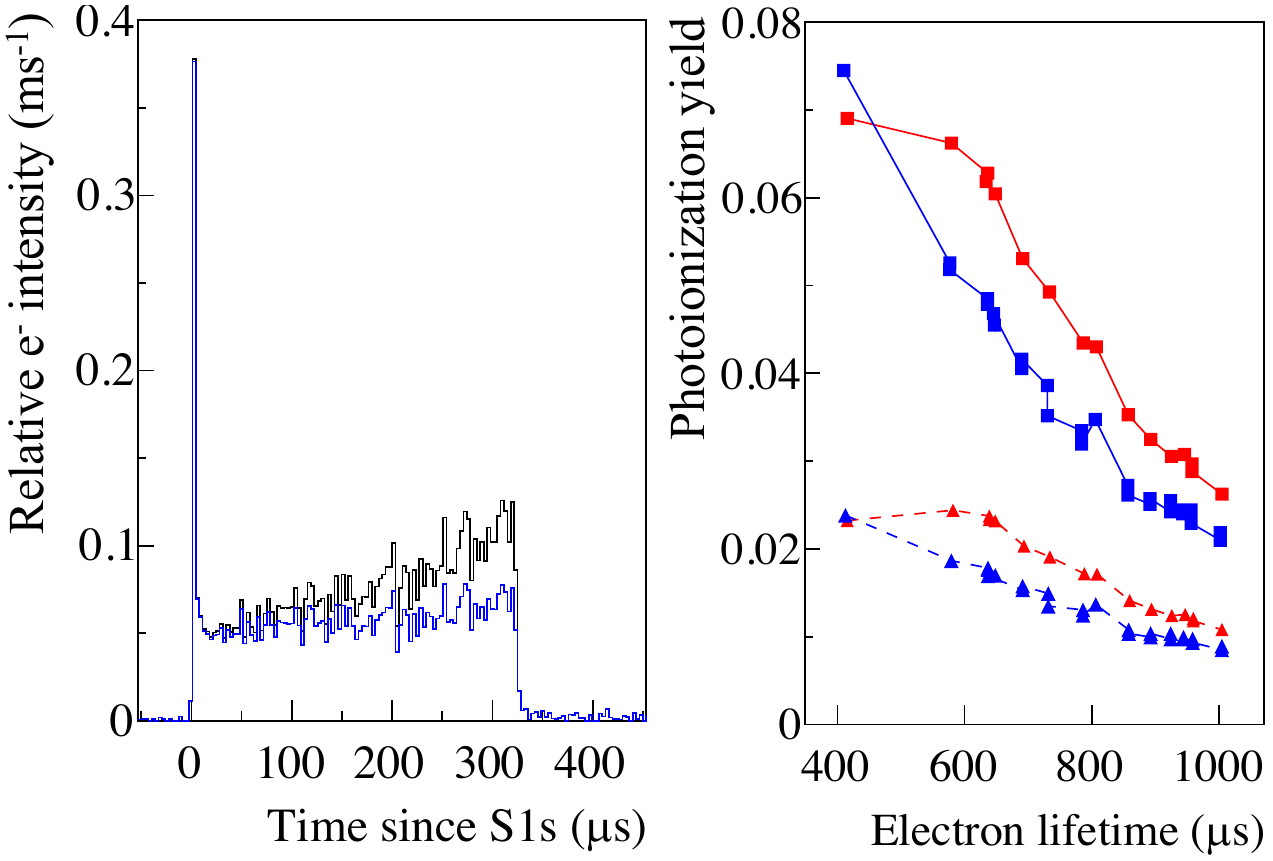}
\caption{{\bf Left:} The relative intensity of electrons after S1-only events, normalized as the electron pulse area divided by the raw S1 area.  
The blue histogram shows the observed values and the black includes a correction for electron loss during drift. 
The peak near $t$$\sim$0 is from photoelectric effect on the extraction (gate) grid. 
{\bf Right: } The photoionization yield (total area of photoionization electrons, divided by the S1 or S2 pulse area) for S1s (blue triangles) and S2s (red squares) as a function of electron lifetime. Both the raw yield values (solid markers) and those corrected with electron loss to impurities and extraction efficiency (hollow markers) are plotted. 
Uncertainties on the data points are estimated to be at the level of 10\% of the corresponding values. 
The lines connecting the data points are not suggestive of any interpolation models, but serve illustrative purposes by keeping the data grouped.}
\label{fig:pirate}
\end{figure}

This immediate electron emission occurs not only following S2s, but also following S1s, consistent with the photoionization explanation. 
Figure~\ref{fig:pirate} (left) shows the time distribution of SEs in the tail of S1s, where the electron intensity is calculated as the ratio of electron pulse area to the S1 area. 
To isolate the features particular to S1 photoionization from S2-related backgrounds, we select only xenon interaction events below  the LUX cathode.
In these events, the nominal S2 pulses are not detected because the ionization electrons are drifted downward by the reversed electric field (referred to as ``S1-only'' events hereafter despite the fact that S1s may produce spurious electron pulses), 
but their positions can be indicated by the dominant S1 signal recorded in the bottom array PMTs. 
In contrast to SEs following S2s, the rate of SEs following S1s increases with time, up to 325~\us. 
This behavior is more apparent after a correction for the electron loss to electronegative impurities during drifts has been applied. 
This difference is explained by the locations of the light source, which is at the bottom of LUX for the S1-only studies and at the top for the S2 studies. 

As shown in Figure~\ref{fig:pirate} (right), the photoionization yields for S1s and S2s are comparable, in the range of 1--10\%, both decreasing at high xenon purities.
The raw yield value is the ratio of the electron pulse area in the 20--320~\us\ delay window following the S1 or S2 pulses to the raw S1 or S2 pulse areas. 
The corrected yield value accounts for the electron losses to impurities and incomplete extraction into the gas (49\% in these data).
The higher yield values for S2s are mostly due to other electron backgrounds following S2s (Sec.~\ref{sec:dse}). 
To reduce this contamination, we only select \kr\ decays within the top 5~cm of liquid xenon, where the additional background is relatively small, in the calculation for S2 photoionization yield. 
We also require no other interaction events in the 10~ms window before the S1 or S2 pulses under study to minimize contamination from previous energy depositions in the detector. 

In the following quantitative discussions, we will focus on the S1-only events, taking advantage of their clean photoionization features. 
Given an average SE size of 25.9~PHE/\el,~\unskip\footnote{The PMT calibration in this work does not compensate for the multi-photoelectron effect~\cite{Faham2015_DPE} because we study both xenon scintillation light and non-xenon photon background (Sec.~\ref{sec:dphe}). Therefore, this SE gain value is slightly different from that used in other LUX analyses.} a photon detection efficiency of 13\% for light emitted below the cathode, and a mean total path length of approximately 2 meters for photons in the LUX liquid xenon predicted by simulations~\cite{LUX2012_LUXSim}, the observed photoionization yield translates to an electron production rate of (5--20)$\times10^{-5}$~\el/$\gamma$/m in LUX. 
Assuming an ionization quantum efficiency of $\mathcal{O}$(1), the corresponding effective photon attenuation length is at the order of 10~kilometers, which is in agreement with a similar calculation in Ref.~\cite{Edwards2009_Thesis}. 
The photoionization process is an insignificant channel for photon extinction in the liquid xenon, 
as also evidenced by the fact that the LUX light collection efficiency is insensitive to the four-fold change of the photoionization electron yield. 

In addition to photoionization in liquid xenon, light from S1s or S2s can also produce photoelectrons on the metal grids that supply electric fields for the TPC operation. 
This is illustrated by the peak structures at $t$$\sim$0 (gate) in Figure~\ref{fig:pi_ratexy} and \ref{fig:pirate} (left) 
and at $t$$\sim$325~\us\ (cathode) in Figure~\ref{fig:pi_ratexy}. 
No cathode photoelectron peaks are observed to follow S1-only events because the photons mostly strike the electrode wires from below; as a result, the liberated electrons will primarily drift downward and cannot be detected. 
The cathode peak is not observed in some low xenon purity data (such as in the red histogram in Figure~\ref{fig:pi_ratexy}) because of the large photoionization population and the strong absorption of electrons from the bottom of LUX by impurities. 
The gate photoelectron peak ($t$$\sim$0) in Figure~\ref{fig:pirate} (left) integrates to $\sim$0.1\% of the detected S1 area. 
With optical simulations using GEANT4~\cite{Agostinelli2003_Geant4, LUX2012_LUXSim}, we estimate that 2.3\% of the below-cathode scintillation light is absorbed by the gate grid, leading to a SS304 stainless steel quantum efficiency of $\sim$4$\times10^{-4}$ for 7~eV photons. 
In this calculation we assume that all photoelectrons from the gate grid surfaces drift upward and can be detected, an approximation supported by electrostatic field simulations using COMSOL~\cite{COMSOL, LUX2017_3DField}. 
Similar calculation for the cathode (SS302) photoelectrons produced by S2 yielded quantum efficiency values of the same order of magnitude, but these carry large uncertainties due to the contamination from additional electron backgrounds and the uncertainty in the fraction of photoelectrons that can be detected. 
This obtained quantum efficiency value is higher than that reported for stainless steel (unspecified grade) by a factor of 2~\cite{Feuerbacher1971_MetalPE}. 
This increase is mostly explained by the reduction of the effective work function of the metal by the electron affinity of liquid xenon~\cite{Tauchert1975_LXeGroundEnergy, Dowell2009_QE}; the contribution from the Schottky effect in reducing the steel work function is subdominant  (10 times smaller than the liquid xenon affinity effect) for the electric fields on the grid surfaces in LUX. 
Other possibilities include differences in the stainless steel grades, the accumulation of positive ions on the grid surfaces~\cite{Malter1936_ThinFilmEmission}, and changes to the electrode surface composition due to collection of positively charged impurities from the liquid. 

\subsection{Photoionization centers in liquid xenon}
\label{sec:pi_xe}

Although the photoionization process in liquid xenon has been discussed by several authors, little is known about the ionization centers other than their connection to electronegative impurities~\cite{ZEPLIN2008_SE, Edwards2009_Thesis, ZEPLIN2011_SE, XENON2014_SE, Huang_PhDThesis}.
An often-discussed candidate for photoionization is negatively charged impurities, such as O$_2^-$, which are formed after electron captures on the electronegative species. 
These negative ions can have a relatively low ionization energy of $<$1~eV, while neutral impurity molecules such as O$_2$, N$_2$ and H$_2$O usually have ionization energies above 10~eV, which appears incompatible with the energy of S1 or S2 photons~\cite{Fujii2015_XeLight, Takahashi1983_ArXe}. 
This hypothesis was tested with LUX data from two different perspectives and was ruled as unlikely. 

\begin{figure}[h!]
\centering
\includegraphics[width=.49\textwidth]{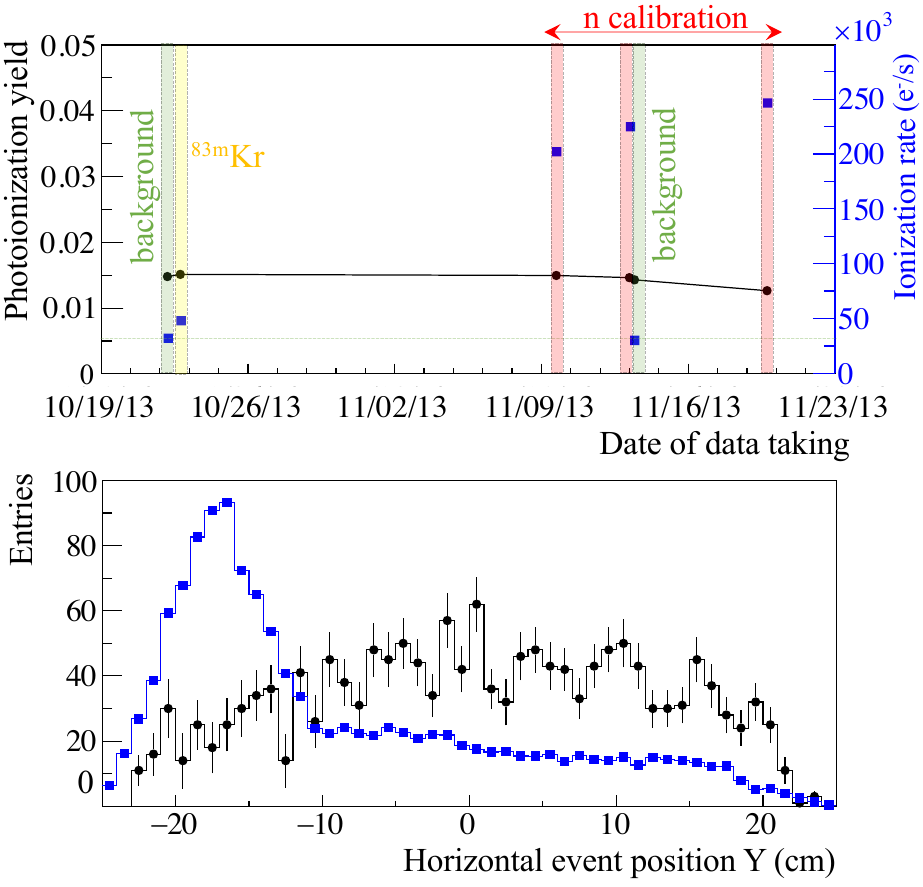}
\caption{{\bf Top:} The S1 photoionization yield values (left scale) and the detected ionization rates (right scale) in LUX under different operating conditions. The photoionization yield values are corrected for evolving electron lifetime using Figure~\ref{fig:pirate} (right). The neutron calibrations were carried out intermittently between November 10th and 21st in 2013 ($>$50\% duty cycle). 
{\bf Bottom: } The Y position distribution for S1 photoionization electrons (black dots) and for total ionization electrons (blue squares, error bars are smaller than the marker sizes due to a down scaling) during neutron calibration. 
The Y direction defined in LUX is approximately parallel to the incident neutron direction and lies in the horizontal plane. }
\label{fig:erate_xy}
\end{figure}

First, if negative ions are responsible for the photoionization emission, the photoionization yield should increase with the negative ion concentration in the liquid. 
The LUX detector was frequently calibrated with internal and external sources, which produced increased rates of ionization.  
During these periods, the rate of negative ion formation through captures of drifting electrons also increased. 
Figure~\ref{fig:erate_xy} (top) shows the varying rates of ionization electrons detected in LUX for background data (30,000~\el/s), \kr\ calibration data (40,000--50,000~\el/s), and deuterium-deuterium neutron calibration data (200,000--250,000~\el/s). 
Despite the significant changes in the expected negative ion formation rate, the S1 photoionization yield remains stable at the 10\% level. 
With a drift speed of 2~mm/s in the LUX liquid xenon~\cite{Hilt1994_IonMobilityLXe}, negative ions such as O$_2^-$ should all migrate to the liquid surface within several minutes; therefore, the negative ion formation rate can be a good indicator for the negative ion concentration in the liquid.~\unskip\footnote{Section~\ref{sec:dse} discusses evidence of negative ions releasing electrons, which also suggests that negative ions may deplete quickly. } 

Second, the three-dimensional positions of the photoionization electrons can be reconstructed and compared to that expected for negative ions. 
Figure~\ref{fig:erate_xy} (bottom) compares the Y positions of S1-induced photoionization electrons (black dots) to that of detected ionization events (weighed by number of electrons) in neutron calibration data (blue squares). 
Although the local energy deposition near the neutron beam entry into the xenon volume ($y$$\sim$$-$20~cm) increased by a factor of 10, 
the horizontal position distribution of photoionization electrons does not exhibit any significant enhancement in this region. 
The Z position of the photoionization electrons also remains consistent with Figure~\ref{fig:pirate} (left) although the additional radiation during neutron calibration was primarily delivered to the upper half of the detector. 
Further, the Z distribution of photoionization electrons as shown in Figure~\ref{fig:pi_ratexy} and Figure~\ref{fig:pirate} (left) can be approximately reproduced with optical simulations~\cite{LUX2012_LUXSim} that assume a homogeneous distribution of ionization centers throughout the liquid xenon. 
It is worth noting that negative ions are expected to have higher concentrations in the upper part of the liquid volume. 
First, all ionization electron clouds are drifted upwards in LUX, which creates a higher electron flux in  the top of the detector and leads to more electron captures in this region.  
Secondly, even negative ions formed in the bottom would drift upward under the influence of the applied electric field. 

In the above discussions, we implicitly assumed that the LUX liquid xenon body was static and charge transport in the detector was solely governed by the electric fields. 
However, convection effects are known to produce a liquid flow speed of $\mathcal{O}$(cm/s) in xenon TPCs~\cite{Malling2014_Thesis, XENON2017_Rn220}, which was measured to be as high as 3~cm/s in certain regions of LUX according to studies using delay coincidence of radioactive-chain decays. 
If this pattern persisted throughout the LUX operation, the convection flow could have reduced the inhomogeneity in the negative ion position distributions, easing the tension from the position comparisons, but the rate argument should still remain valid. 
In this scenario, the concentration of negative ions in LUX is no longer determined by the prompt production rate of negative ions but its integrated history. 
In LUX the xenon circulation turn-around time was 1--3 days, 
and during this process all ions should lose their charge states. 
Therefore, with approximately 10 days of neutron calibration ($>$50\% duty cycle), the negative ion concentration should reach an equilibrium with the increased radiation level, 
and yet no proportional increase of the photoionization yield was observed. 

Based on these observations, we rule out negative ions from dominating the photoionization process, and, instead, propose that some neutral impurities must play a leading role. 
However, given the strong correlation between the photoionization yield and the electron lifetime, this neutral impurity should be present in proportion with the electronegative species. 
Therefore, the magnitude of this photoionization background can be used as a liquid xenon purity monitor in lieu of dedicated source calibrations, as successfully demonstrated by ZEPLIN-III~\cite{ZEPLIN2011_SE}, LUX~\cite{Huang_PhDThesis} and XENON100~\cite{XENON2014_SE}. 
In addition, the single electron-ion pair produced by the photoionization effect can be easily separated by weak electric fields, and thus the electron collection efficiency is insensitive to modest changes in the local electric field, a behavior observed for low ionization density interactions in liquid xenon~\cite{Aprile2017_ERField, Xu2019_EEE}. 
Combining with the fact that SE pulses are unlikely to suffer signal distortions, the photoionization electrons can provide a robust charge collection efficiency calibration even in non-uniform electric fields. 

\section{Clustered electron emission}
\label{sec:eburp}

\begin{figure}[b!]
\centering
\includegraphics[width=.4\textwidth]{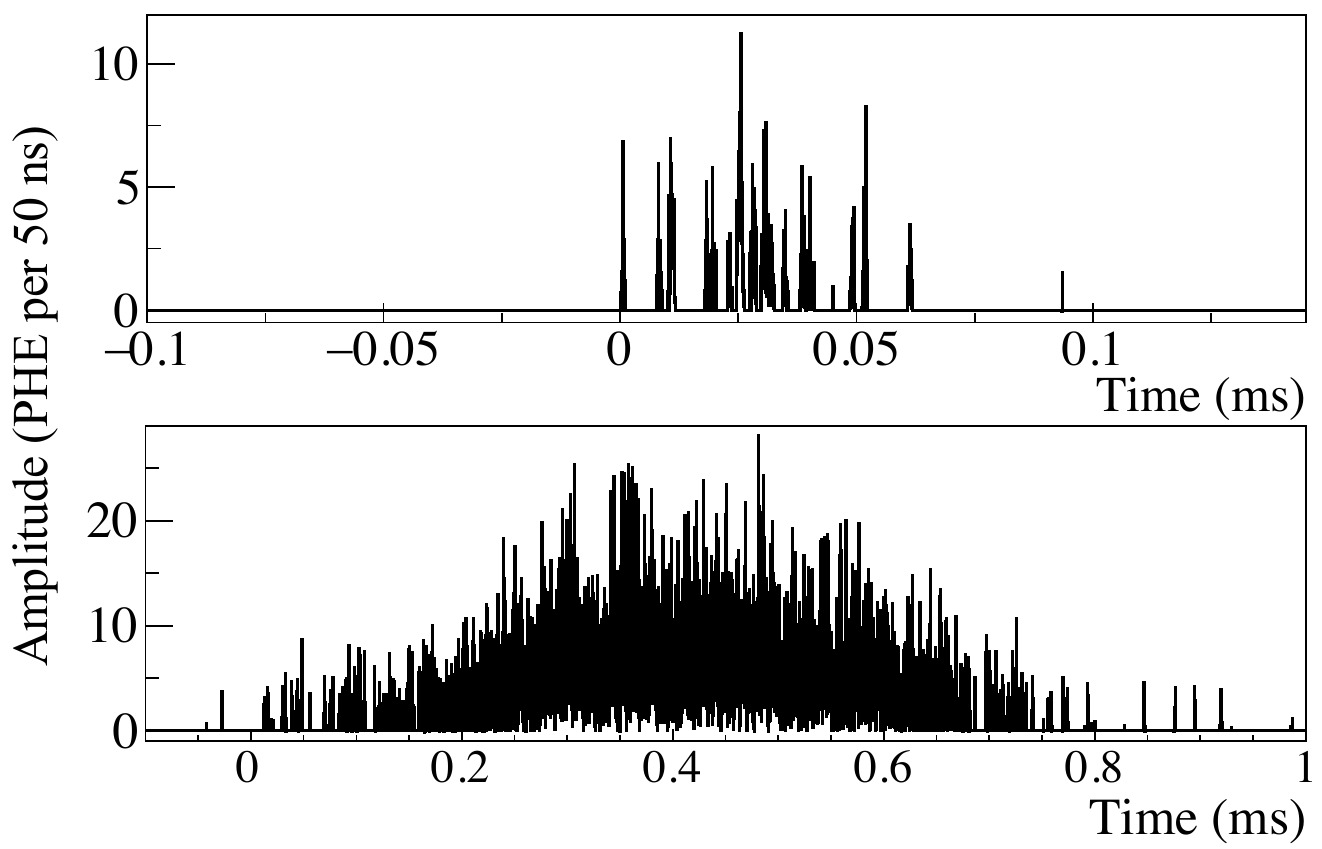}
\caption{Example waveforms of clustered electron emission (``e-bursts'') observed in LUX. 
E-bursts do not contain S1 pulses or S2 pulses other than clustered SEs. 
The e-burst in the top contains two dozen SEs and the bottom one contains more than two thousand overlapping SEs. 
The waveform samples are grouped into time bins of 50~ns to preserve the SE pulse structure (1--2~\us) but differentiate their amplitudes (5--10) from that of PHE pulses ($\sim$1).
}
\label{fig:eburp_wfm}
\end{figure}

The most prominent electron emission pathology observed in LUX is clustered electron emission, which is already illustrated in Figure~\ref{fig:wfm} (top right inset), and two more example waveforms are shown in Figure~\ref{fig:eburp_wfm}. 
Such a cluster contains a number of pulses, each of which is consistent with the electroluminescence light produced by extracted electrons.
However, contrary to normal S2s that correspond to a cloud of electrons being detected within a few microseconds, these electron clusters consist primarily of individual SE pulses that are spread over a larger time window of 10~\us--1~ms. 
The clusters are usually preceded and succeeded by quiet periods, and thus cannot be attributed to statistical fluctuation of SE rates. 
Despite large variations in size, the clusters exhibit similar timing structures, with the rate of electron pulses rising slowly in the beginning and falling in the end. 
Similar pathology events were also reported in XENON10~\cite{Sorensen2008_PhDThesis}. 
In LUX, these clusters  (referred to as ``e-bursts'' hereafter) are observed to follow high-energy particle interactions in the liquid xenon at the millisecond to tens-of-milliseconds scale, and most of the SEs in a clusters share the same X-Y location as the preceding S2. 
Because multiple electrons may be emitted around the same time, small e-bursts can pose a significant background for dark matter searches using ionization-only events. 

\begin{figure}[h!]
\centering
\includegraphics[width=.49\textwidth]{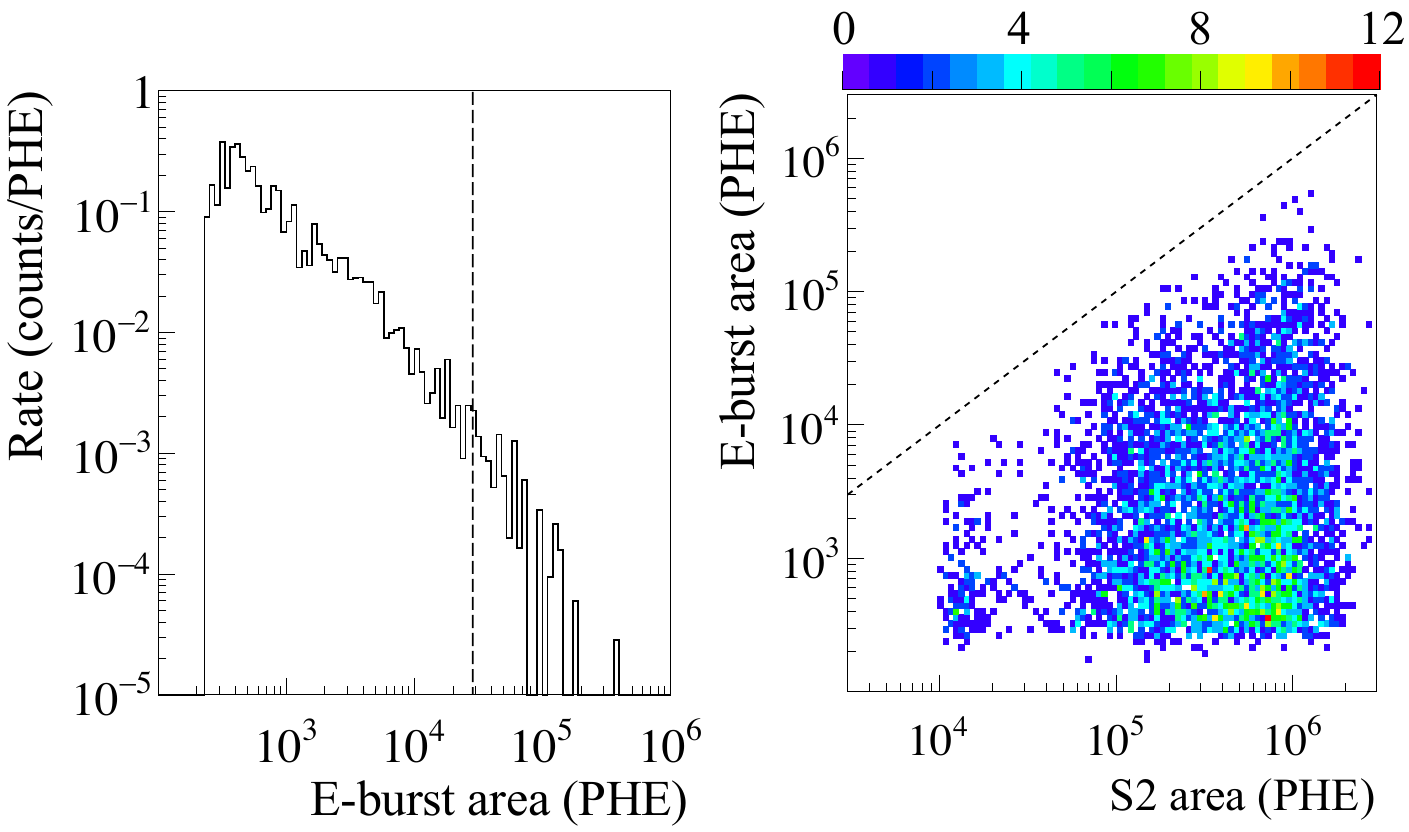}
\caption{{\bf Left:} The energy spectrum of e-burst clusters observed to follow large S2s (6$\times$10$^5$--8$\times$10$^5$~PHE). 
The vertical dashed line separates the top 1\% e-burst area from the rest of 99\%, corresponding to a $R_{99\%}$ value of 0.04 in this example (92\% of large S2s studied in this data set do not produce e-bursts above threshold and are off the log axis.).  See text for more explanations on the $R_{99\%}$ calculation. 
{\bf Right: } The summed e-burst area within a 50~ms window following a large S2 as a function of the raw S2 area. The dashed line corresponds to equal e-burst area to the S2 area. 
For reference, the average SE area is 25.9~PHE.}
\label{fig:eburp_size}
\end{figure}

An e-burst may contain fewer than ten electrons, or over tens of thousands of electrons in some cases. 
Figure~\ref{fig:eburp_size} (left) shows the size distribution of e-bursts identified following large S2s in a WIMP-search data set. 
The spectrum is largely featureless, and decreases monotonically with the e-burst area. 
The low-energy cutoff below 300~PHE is due to the 10-pulse threshold used to tag e-bursts in this work; below this threshold smaller e-burst clusters should exist but are difficult to distinguish from random pileups of SEs in high electron rate periods. 
The upper bound of e-burst area distributions, however, is observed to correlate with the area of preceding S2s, as illustrated in Figure~\ref{fig:eburp_size} (right). 
In the data investigated, the maximum e-burst area values are typically 10--50\% of that of preceding S2s; in these extreme cases the e-burst clusters contain a number of electrons close to half of that in the preceding S2. 
Approximately 90\% of large S2s do not produce e-bursts above the analysis threshold, but around one percent of them are followed by two or more such clusters. 
When evaluating the e-burst area for these events, we use the summed e-burst area in the 50~ms window following the S2s. 

\begin{figure}[h!]
\centering
\includegraphics[width=.45\textwidth]{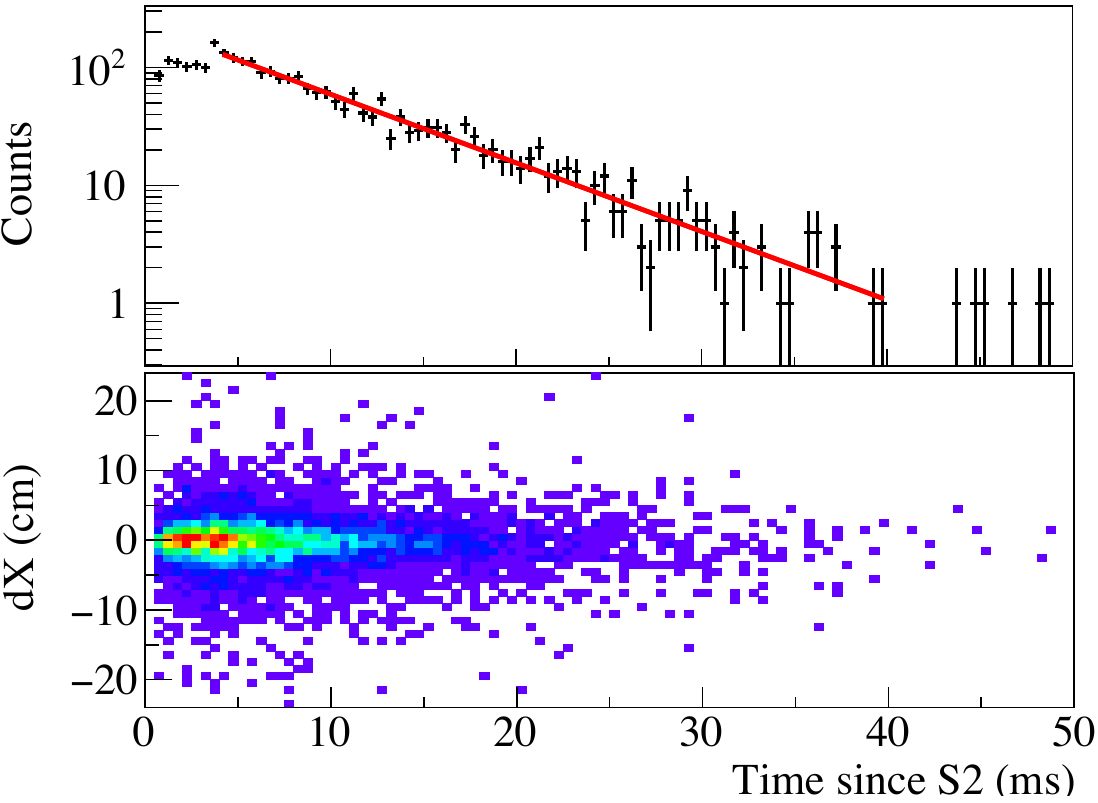}
\caption{{\bf Top:} Time delay of e-bursts from their preceding S2s in one WIMP search data set. 
The red curve shows an exponential fit to the distribution, which yields an exponential slope of 7.5$\pm$0.3~ms; 
as discussed later, this slope has a significant dependence on event positions and other parameters. 
{\bf Bottom:} The X position difference between e-bursts and preceding S2s, plotted as a function of the time delay. 
In these analyses, we require that there be no additional large S2 pulses in the period of $[-30, +50]$~ms relative to the S2s of interest. }
\label{fig:eburp_dtdx}
\end{figure}

The distribution of time delay between e-bursts and their preceding S2s can be described by a single exponential, with a decay constant (characteristic delay time) of $<$10~ms, as illustrated in Figure~\ref{fig:eburp_dtdx} (top). 
We comment that due to the Poisson nature of particle interactions, the time separation between an uncorrelated process in LUX and the preceding S2 pulse will also follow an exponential distribution, the decay constant of which (250--300~ms) is governed by the LUX event rate (3--4~CPS). 
In this analysis, by requiring only one large S2 pulse in the time window under study, we have removed this underlying exponential component so we can focus on real time correlations. 
The exponential nature of the e-burst time delay is consistent with a model where these electrons are supplied from a reservoir that is filled around the time of the large S2s and is continuously drained at a fixed rate over time. 
Competing processes for draining this reservoir should exist; otherwise, the summed e-burst area should be more directly correlated with the preceding S2 area. 
Figure~\ref{fig:eburp_dtdx} (bottom) also shows the position difference between e-bursts and their preceding S2s as a function of the time delay. 
The electron pulses in a cluster usually share the same X-Y positions, which also coincide with that of preceding S2s. 
This position correlation does not appear to weaken over time up to 50~ms. 

Based on the observed energy, time and position correlations between e-bursts and their preceding S2s, we propose that this clustered electron emission results from one of the direct electron sources discussed in Sec.~\ref{sec:source}, including unextracted electrons and impurity-captured electrons. 
Hypotheses involving primary positive ion clouds are disfavored because of their small mobility in liquid xenon. 
With an estimated drift velocity of 8~mm/s at the LUX drift field~\cite{Hilt1994_IonMobilityLXe}, it would take minutes for a positive ion cloud to reach the cathode grids, and thus cannot explain the immediate emission of e-bursts (within ms after S2s). 
Similarly we can rule out other processes that require ion drift, such as neutralization with negative charges, from playing a major role in the e-burst emission due to the much longer expected timescale (seconds or longer). 
The ions that may be produced in the high-field gas regions near LUX anode wires could travel down to the liquid surface within a few milliseconds~\cite{Neves2007_XeGasMobility}, 
but due to the fixed traveling distance, these ion activities should occur with a constant time delay from the S2s, rather than producing an exponentially decaying background. 

To further test these hypotheses, we quantitatively describe the e-bursts  under different experimental conditions using their size and rate of decay over time since S2s. 
The size is characterized using the maximum e-burst area normalized to the observed area of the preceding S2; 
the maximum e-burst area is approximated with the 99-percentile value, and the obtained ratio is referred to as $R_{99\%}$ hereafter. 
When calculating $R_{99\%}$, we also include large S2s that are not followed by identifiable e-bursts, so this calculation is not biased by the inefficiency in tagging small e-bursts. 
An illustration of the 99-percentile e-burst area determination and the $R_{99\%}$ calculation can be found in Figure~\ref{fig:eburp_size} (left).

\begin{figure}[h!]
\centering
\includegraphics[width=.48\textwidth]{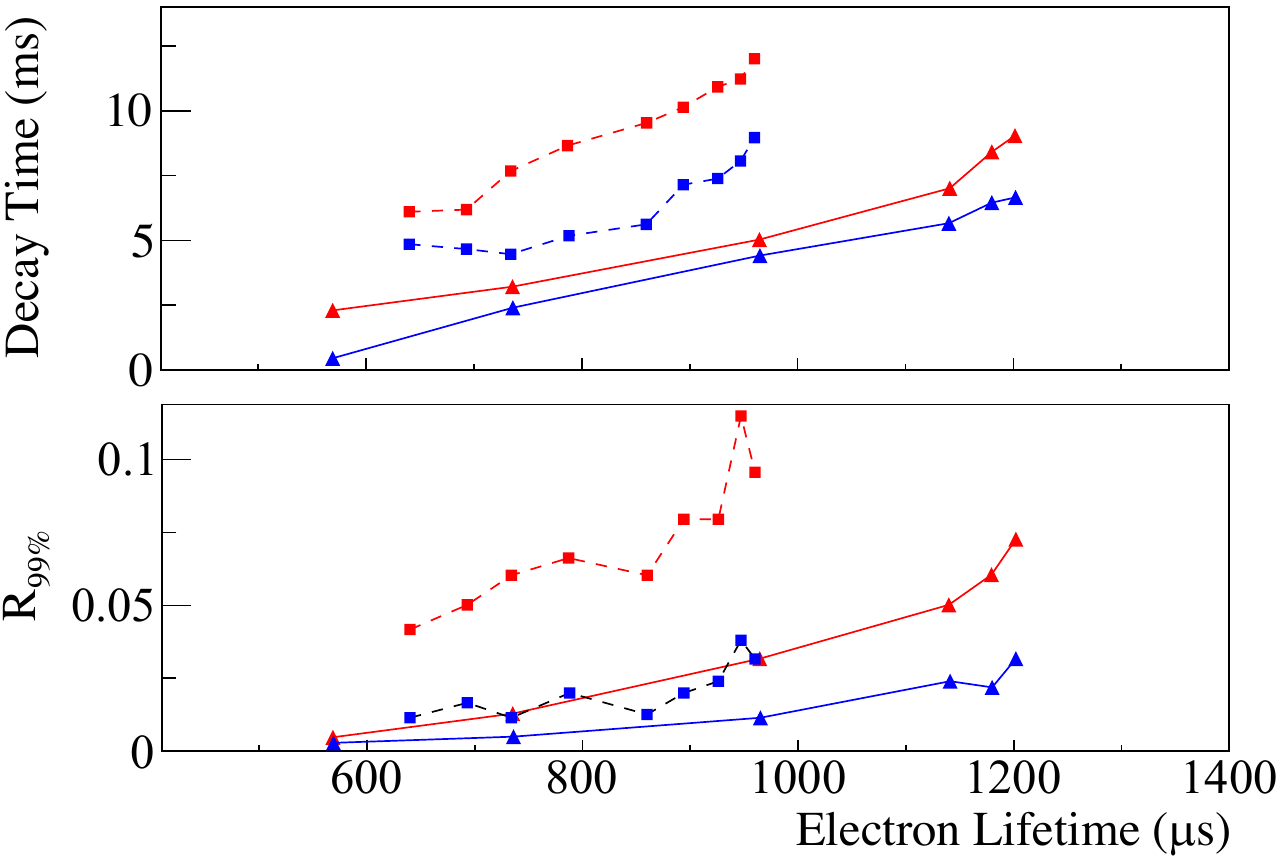}
\caption{{\bf Top: } The observed decay time of e-burst clusters following interactions in the top (red) and bottom (blue) 5~cm of LUX liquid xenon in WS2013 (squares) and WS2014--16 (triangles). 
{\bf Bottom: } The $R_{99\%}$ parameter (defined in the text) as a function of electron lifetime (same legend notation as the top subfigure); the purity and position dependences would be amplified if we calculate $R_{99\%}$ with the electron lifetime-corrected S2 areas.  
Uncertainties on the data points are estimated to be at the level of 10\% of the corresponding values. 
The lines connecting the data points keep the data grouped for illustrative purposes.
}
\label{fig:eburp_rate}
\end{figure}

Figure~\ref{fig:eburp_rate} shows the characteristic decay time (exponential slop illustrated in the top of Figure~\ref{fig:eburp_dtdx}) of e-bursts (top) and their $R_{99\%}$ values (bottom) at different liquid xenon purity (measured as electron lifetime) levels. 
The results are separately plotted for interactions in the top (5~cm below the liquid surface) and bottom (5~cm above the cathode grid) of LUX, in WS2013 (49\% electron extraction) and in WS2014-16 (73\% electron extraction).
Generally speaking, the e-burst emission becomes stronger and also lasts for longer as the liquid xenon purity improves. 
Despite sharing similar purity dependence, particle interactions in different regions of the detector lead to different e-burst behaviors. 
Compared to e-bursts following interactions near the top of LUX (red markers in Figure~\ref{fig:eburp_rate}), the decay of e-bursts following bottom-originating S2s (blue markers in Figure~\ref{fig:eburp_rate}) occurs faster 
and these e-bursts are on average smaller in size. 
This top-bottom disparity disfavors the impurity-captured electrons as an explanation for the e-burst emission, because bottom-originating ionization events produce more negatively charged impurities due to the longer electron drift, a trend contradicting the observation. 
Rather, the weakening of e-burst emission with higher concentration of impurities in liquid xenon indicates that impurities are a competitor to e-burst productions. 
The same difficulty applies to other processes involving impurity-captured electrons such as the combination of positive and negative ion clouds. 

Unextracted electrons trapped under the liquid surface provide a strongly favored electron source for e-bursts. 
In LUX WS2013 data, with an extraction electric field of 3.5~kV/cm below the liquid surface, approximately 49\% of the ionization electrons arriving at the liquid surface were extracted into the gas while the rest became trapped. 
If some trapped electrons can emerge into the gas together under certain excitation, the electron emission will exhibit energy, time and position correlations with preceding S2s similar to that observed for e-bursts. 
In addition, while the electrons are trapped under the liquid surface, some will be captured by impurities and thus cannot contribute to e-bursts, which naturally explains the the anti-correlation between e-bursts and the impurity levels in the liquid xenon. 
This alternative outcome for the surface-trapped electrons also explains why only the maximum e-burst area is correlated to the total number of trapped electrons (proportional to the observed S2), as illustrated in Fig.~\ref{fig:eburp_size}.
Further, as shown in Figure~\ref{fig:eburp_rate}, an increased electron extraction efficiency in WS2014--16 (4.2~kV/cm extraction field, 73\% efficiency) leads to fewer trapped electrons, and consequently the e-burst emission dies out more quickly and the e-burst size becomes smaller.~\unskip\footnote{Due to the evolving electric field distortion in WS2014--16, the calculation of drifting electron lifetime in LUX liquid using \kr\ decays carries an additional source of uncertainty. However, this statement should hold as long as the evaluated lifetime values are accurate within a factor of 2.}

However, this explanation faces two difficulties: 1) how can quasi-free electrons preserve their X-Y positions under the liquid surface for tens of milliseconds, and 2) why is the electron lifetime at the liquid surface 5--10 times larger than that in the liquid bulk? 
These two challenges may be simultaneously addressed if a deformation of the liquid surface occurs where the electrons are trapped. 
The presence of dense electric charge under the liquid surface in a strong electric field can raise the local liquid level microscopically; 
this local liquid level deformation, together with the vertical electric field, can function as a physical trap and preserve the X-Y position of trapped electrons for a long time. 
At the same time, being trapped in a small volume could limit the exposure of these electrons to impurity molecules and alter the velocity-dependent capture cross section~\cite{Bakale1976_EfieldAttach}, so that  the observed electron lifetime is significantly increased. 
Moreover, if this hypothesis is correct, higher density electron clouds, such as those from the top-originating ionization events where transverse diffusion~\cite{EXO2017_EDrift} is less significant, 
can produce stronger traps and thus explain the observed top-bottom disparity. 

Regarding the underlying mechanism that may excite unextracted electrons to be emitted from the surface in clusters, one possibility is the movement of the liquid xenon surface, such as capillary waves, which may be generated by the xenon flow, formation of bubbles in liquid, etc. 
The e-burst widths of $\lesssim$1~ms correspond to an excitation frequency of kHz, which matches that of capillary waves for liquid xenon estimated from its surface tension~\cite{Smith1967_XeSurfaceTension}. 
Amid upward oscillations of the liquid surface, the trapped electrons gain kinetic energy from the strong electric field and at the same time dissipate part of the gained energy to xenon atoms through collisions. 
This process is in direct analogy with the heating of primary electron clouds when they first reach the liquid surface, and if sufficient energy is gained the electrons can be extracted into the gas~\cite{Gushchin1979_EEE}. 
In addition, the strongest e-burst emission region in LUX ($X$$\sim$0, $Y$$\sim$$-23.5$~cm) coincides with the location of the largest detected S2 signal areas for mono-energetic \kr\ delays in the liquid~\cite{LUX2017_Kr83m}, which also suggests unusual activities on the liquid xenon surface in this region. 
Other forms of delayed emission may also occur for these trapped electrons, such as thermionic emission~\cite{Sorensen2017_EEE}, but no significant evidence for a fast emission component as reported in Ref.~\cite{SorensenElectronBG_2018} is observed in this work. 

\section{Delayed background emission}
\label{sec:delayed}

If one excludes the photoionization electrons and the e-burst clusters, the remaining background in LUX mostly consists of individual SE and PHE pulses, as illustrated in Figure~\ref{fig:wfm}. 
The rates of SEs and PHEs both sharply increase after large S2s, and then decrease slowly with time up to the longest timescale (1 second) that can be studied with LUX data. 
In this section, we will characterize these backgrounds and discuss their possible origins. 

\subsection{Individual electron background}
\label{sec:dse}

Individual SE background is observed in the LUX detector at timescales from a few microseconds following a S2 to seconds later.  
Due to this large time span, we study the SEs in two analysis windows: a) 0--3~ms and b) 3--1000~ms following S2s. 
For the short window (0--3~ms) analysis, we select \kr\ calibration events to obtain high statistics, 
and for the long window (3--1000~ms), we use low trigger rate WIMP-search data that have large time intervals between events. 
In each scenario, we select the events with no other particle interaction in the event time window. 
As explained in Sec.~\ref{sec:eburp}, this criterion is necessary to get rid of the exponential feature that arises from the Poisson nature of particle interactions in LUX. 
In addition, we require no other energy depositions in the detector within the preceding 100~ms of the S2 under study so that contamination from earlier events is reduced. 
Electrons in e-burst clusters, as described in Sec.~\ref{sec:eburp}, are excluded from this analysis. 

\begin{figure}[h!]
\centering
\includegraphics[width=.49\textwidth]{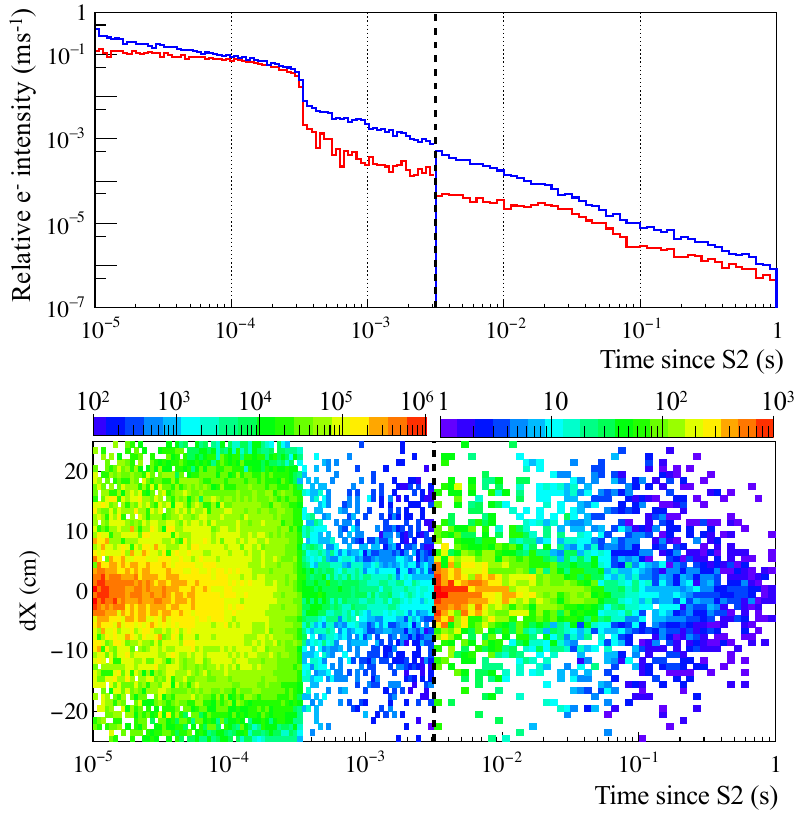}
\caption{{\bf Top:} The relative electron intensity (defined in text) as a function of delayed emission time. 
The red (blue) spectrum is for SEs following energy depositions in the top (bottom) 5~cm of the active LUX liquid xenon. 
{\bf Bottom: } The difference of X position between SEs and preceding S2s as a function of delay time. 
The S2s are chosen to be in the center 5~cm radius of LUX so the $dX$ distribution is not biased due to the finite size of LUX for $dX<$18~cm. 
Different sets of data are used to generate the plots $<$3~ms and $>$3~ms, leading to a small discontinuity. This transition is indicated by the vertical dashed line. 
}
\label{fig:se_ratedx}
\end{figure}

The rate of SEs following a large S2 is approximately proportional to the S2 area, 
so in the following analysis we define the relative SE intensity as the observed SE areas divided by the raw S2 area. 
Figure~\ref{fig:se_ratedx} (top) shows the relative SE intensity as a function of the emission time following S2s originating from the top (red) and the bottom (blue) of LUX. 
Both spectra show similar time dependences, including a clear photoionization cutoff at the maximum S2 drift time of 325~\us\ (Sec.~\ref{sec:pi}) and a long power law-like tail up to 1~s. 
The approximate power-law form is between t$^{-1.1}$ and t$^{-1.0}$.
At all delay time values, the SE rate is higher following interactions in the bottom of LUX compared to that in the top, 
and this position dependence remains if we calculate the SE intensity using the electron lifetime-corrected S2 area. 
In addition, the majority of this electron population exhibits a strong X-Y position correlation with the preceding S2s, as illustrated in the bottom of Figure~\ref{fig:se_ratedx}. 
These observations rule out cascade photoionization (S2 - photoionization - SEs - photoionization, and so forth) as a significant contributor to this tail. 
Also the relatively low photoionization yield means that the amplitude of this cascade should decrease by 1--2 orders of magnitude for every 325~\us\ and will become insignificant after 1~ms. 

\begin{figure}[h!]
\centering
\includegraphics[width=.42\textwidth]{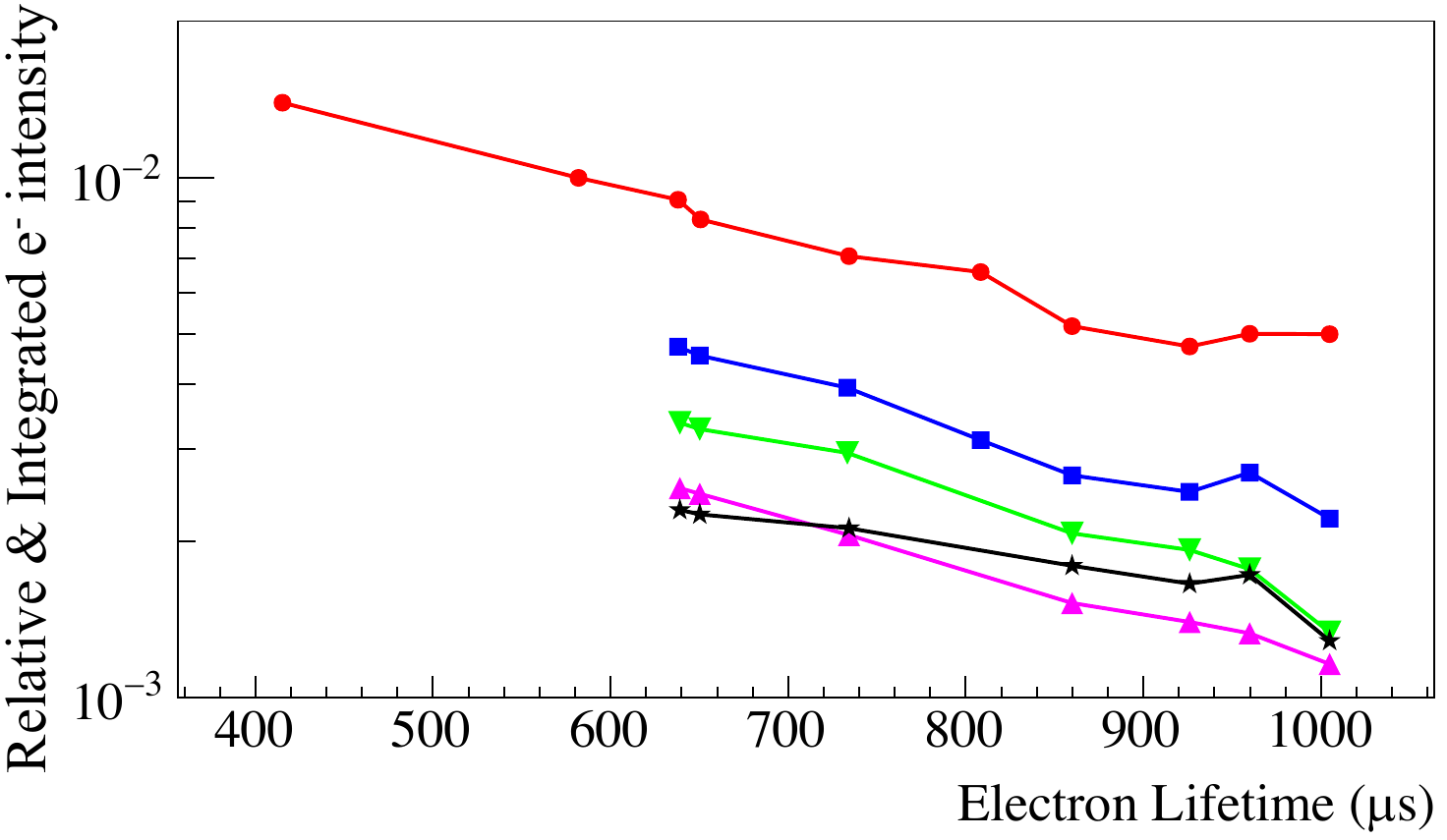}
\caption{The relative SE intensities integrated over different delay time windows after the S2: 10--500~\us\ (red dots), 500~\us--3~ms (blue squares), 3--10~ms (green downward triangles), 10--100~ms (purple upward triangles), 100--1000~ms (black stars), 
as functions of liquid xenon purity levels associated with different periods of detector operation. 
Uncertainties on the data points are estimated to be at the level of 10\% of the corresponding values. 
Here we select only SEs after particle interactions in the bottom 5~cm of LUX because they produce the most delayed SEs. 
Data points for the delay window of 10--500~\us\ have the S2 photoionization contributions subtracted already. 
The lines connecting the data points only serve illustrative purposes by keeping the data grouped.
}
\label{fig:serate_purity}
\end{figure}

We further studied the dependence of this background intensity on the liquid xenon purity, as summarized in Fig~\ref{fig:serate_purity}. 
In all delay time windows studied, the electron intensity decreases as the electron lifetime increases. 
This purity dependence remains qualitatively the same for SE intensities calculated using the electron-lifetime corrected S2 area.  
The dependence of SE intensities on the Z position of preceding interactions (Fig.~\ref{fig:se_ratedx}, top) and on the xenon purity (Fig.~\ref{fig:serate_purity}), together with the X-Y position correlation (Fig.~\ref{fig:se_ratedx}, bottom), indicate that they originate from electrons captured by impurities~\cite{SorensenElectronBG_2018}. 
Ionization electrons produced by particle interactions near the bottom of LUX need to drift across large distances in liquid xenon and thus will leave behind more negatively charged impurities. 
Due to their low mobility and diffusivity~\cite{Hilt1994_IonMobilityLXe}, negatively charged impurities can preserve their X-Y positions for seconds or longer. 
So, if they release the electrons under certain excitations~\cite{DS50_2018_S2Only}, the resulting electrons can exhibit the observed energy and position correlations with the S2s. 
In addition, as discussed in Sec.~\ref{sec:eburp}, some unextracted electrons under the liquid surface are also captured by impurities, which can explain the non-zero delayed SE emission following particle interactions near the top of LUX (Fig.~\ref{fig:se_ratedx}, top). 

As illustrated in Fig.~\ref{fig:se_ratedx} (bottom), around 10\% of the delayed SEs do not share X-Y positions with preceding S2s. 
This population includes photoionization electrons generated by EL photons from prior SEs and also delayed electron emission by negative ions produced during previous interactions in the detector. 
The relatively low rate of this population suggests that the responsible negative ions are extinguished quickly in liquid xenon, which may occur by releasing the electrons, neutralizing with positive ions, or through other processes such as drifting to the liquid surface and spilling into the weir, supporting the hypothesis assumed in Sec.~\ref{sec:pi}. 

As for the mechanism leading to the power law-like time dependence of the delayed SEs, it has been reported that electron-ion recombination in liquid xenon~\cite{Kubota1979_ArKrXeRecombination}, bi-exciton quenching in liquid scintillators~\cite{King1966_LSTime}, and the pressure reduction in an elementray vacuum system~\cite{Edwards1977_VacuumChamberPowerLaw} can all produce time dependences dominated by power-law features. 
If the behavior of negative ions in liquid xenon is subject to similar dynamics, their rate can exhibit a power-law decrease over time, 
and processes such as collisional ionization~\cite{SorensenElectronBG_2018} could lead to SE emission as observed in this work. 
Another possibility for negatively charged impurities to release electrons is photoionization by background optical or infrared photons in LUX, to be elaborated in Sec.~\ref{sec:dphe}. 

\subsection{Photoelectron background}
\label{sec:dphe}

\begin{figure}[h!]
\centering
\includegraphics[width=.49\textwidth]{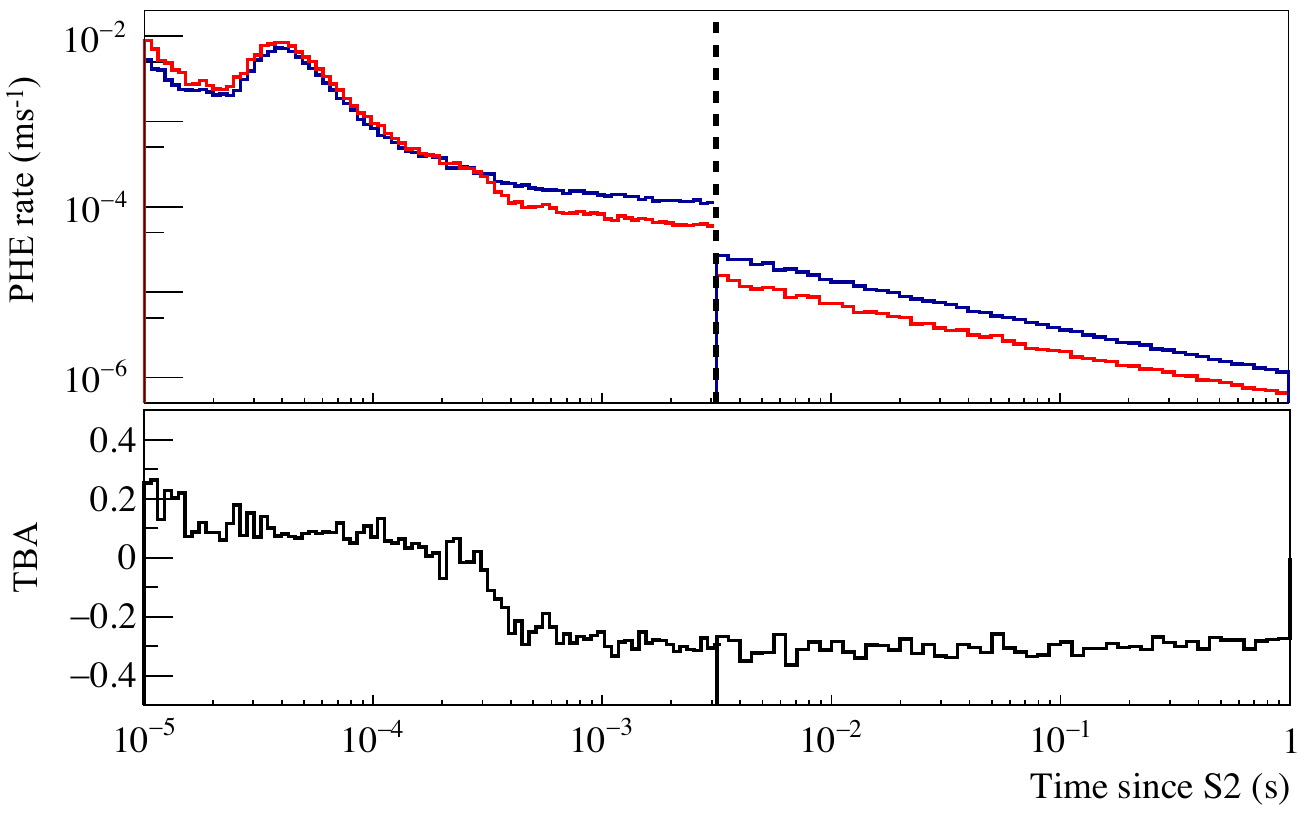}
\caption{{\bf Top:} The PHE rate observed in the top (red) and bottom (blue) PMT array of LUX as a function of time since the preceding S2s. 
The discontinuity at 3~ms is a result of the two analysis time windows explained in Sec.~\ref{sec:dse}.
{\bf Bottom: } The top-bottom asymmetry (T-BA) of the PHE background as a function of time since S2s. In LUX, S1 PMT signals have a T-BA value of $-0.3$ (near liquid surface) to $-0.8$ (near cathode), and S2 has an average T-BA value of 0.16. }
\label{fig:spe_ratetba}
\end{figure}

As illustrated in Figure~\ref{fig:wfm} and discussed in \cite{LUX2020_DPE}, an increase of the PHE rate is observed in LUX after particle interactions in the TPC. 
These PHEs do not exhibit a double-photoelectron emission feature~\cite{Faham2015_DPE}, suggesting that they have longer wavelengths than the vacuum ultraviolet (VUV) xenon photons associated with the regular S1 and S2 emission processes. 
Figure~\ref{fig:spe_ratetba} (top) shows the PHE rate in the LUX detector following high-energy S2s. 
This plot is produced similarly to Figure~\ref{fig:se_ratedx} (top), but we further require the PHE pulses are not in the immediate vicinity of (1~\us\ before and after) SEs or S2s to exclude misidentified photons that are part of SE or S2 pulses.
Similar to SEs, the PHE rate exhibits a gradual power law-like decrease over time. 
Beyond 325~\us\ past the S2 time, the ratio of the PHE rates in the top PMT array to that in the bottom array remains approximately 1:2, leading to a top-bottom asymmetry (T-BA) value $(A_T-A_B)/(A_T+A_B)=-0.3$ (Figure~\ref{fig:spe_ratetba} bottom). 
This observation disfavors the explanations of these PHEs as thermionic dark noise or other PMT instrumental effects due to the equal PMT numbers in the two arrays and the higher VUV photon rate detected by the top array from S2s. 
If we assume that these PHEs result from photons emitted from a single location in LUX, the T-BA value of $-0.3$ indicates that the light source may reside right below the liquid surface or near the gate. 
An alternative explanation of the PHEs is the fluorescence of PTFE reflectors~\cite{Shaw2007_PTFE} surrounding the whole active xenon volume. 
This hypothesis can produce a similar T-BA value to that observed if the PTFE fluorescence has a position-dependent emission strength proportional to the number of VUV photons absorbed at the same location, and also has similar optical transportation properties to those of xenon VUV light.

In contrast to the  background electrons, the PHE rate is relatively insensitive to the xenon purity or the position of the preceding interaction. %
The PHE rate also decreases more slowly with time (between $t^{-0.6}$ and $t^{-0.5}$) than the SE rate ($t^{-1.1}$--$t^{-1.0}$). 
Therefore, we rule out the possibility of these photons being emitted as a by-product of the electron emission process but, instead, they may be a triggering mechanism for the photoionization electron emission by negatively charged impurities (Sec.~\ref{sec:dse}). 
Due to the unknown optical properties and detection efficiency of the photons that may produce this PHE background, we cannot directly compare the electron-to-photon rate in this section to the VUV photoionization yield obtained in Sec.~\ref{sec:pi}. 
A full characterization and explanation of this PHE background is beyond the scope of this work. 

\section{Multiple-electron background in LUX}
\label{sec:me}

The majority of electron background discussed thus far consists of SEs,  
while current searches for low-energy dark matter interactions using ionization-only signals are more susceptible to multiple electron (ME) backgrounds.  
During high electron rate periods shortly after large S2s, random coincidence of SEs can produce pile-up ME pulses, 
and phenomena such as e-bursts can also induce ME emission. 
These ME events in the aftermath of S2s dominate the raw electron spectrum observed in LUX, and prevent genuine ME events from being studied. 
Fortunately, such accidental ME pulses, similar to SEs, have strong time correlation with previous energy depositions in the detector and 
can be substantially suppressed through a veto cut after high-energy events. 

\begin{figure}[h!]
\centering
\includegraphics[width=.49\textwidth]{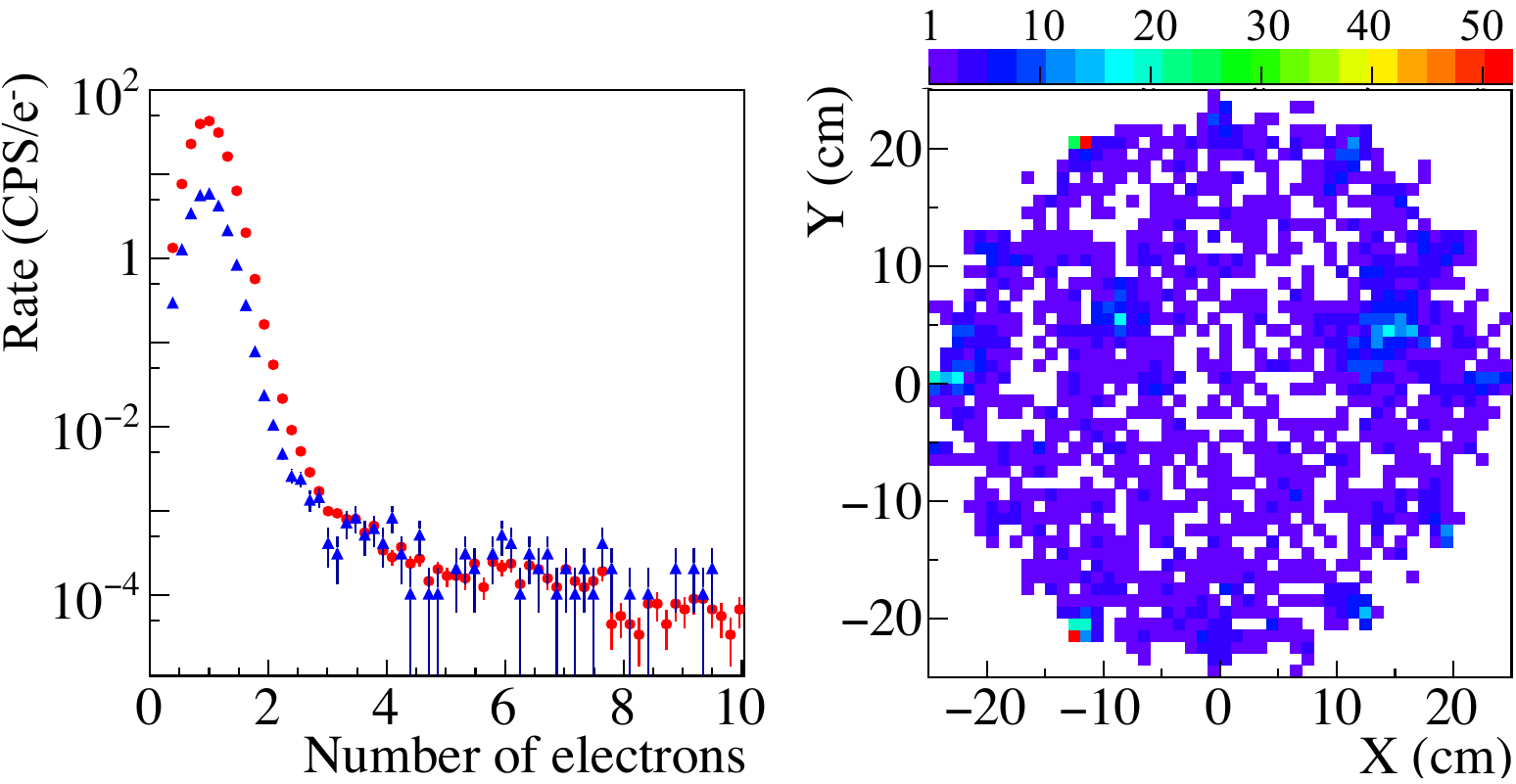}
\caption{{\bf Left:} Residual electron event rate (live-time corrected) in 10 days of LUX dark matter search data with a conservative (red) and an aggressive veto cut (blue) after large S2s; intense periods of hot-spot electron emission (shown in the right figure) have been excluded; 
{\bf Right: } The X-Y position distribution of residual electron pulses above 2.5 extracted electrons. The hot spots in the inner volume are attributed to grid electron emission. }
\label{fig:serate_vetoed}
\end{figure}

Figure~\ref{fig:serate_vetoed} (left) shows the residual electron event spectra in ten days of LUX WIMP-search data using two veto algorithms: 
the first method (red histogram) uses a simple 50~ms veto cut after each identified interaction of 1~keV or above; 
the second (blue histogram) employs an aggressive veto cut of 50--5000~ms, the value of which increases with the S2 area and drift time to compensate for the higher SE rates. 
A minimum veto window of 50~ms is applied to both scenarios to get rid of pile-up electron pulses and small e-bursts. 
The live-time loss due to the veto cut is 13\% for the first method and 90\% for the second. 
In addition to the veto cut, we require no other S2 or SE pulses, or S1 pulses of greater than 3~PHE, in the same event window (350~\us) as the pulses of interest. 
Additional efficiency losses due to this cut are estimated to be less than 5\%. 
The data selected have relatively high and stable liquid xenon purity (900--950~\us\ electron lifetime), which, as discussed in Sec.~\ref{sec:dse}, leads to less delayed electron emission. 

The residual electron pulse rates after the veto cuts are substantially reduced compared to that without the veto ($\mathcal{O}$(10$^3$~CPS)). 
With a 50~ms veto, the live-time corrected rate of SEs is found to be 26~CPS, and it decreases to 4~CPS with the aggressive veto cut in the second method. 
Because of the slow decay (Figure~\ref{fig:se_ratedx}), the post-veto SE rate is expected to continue decreasing for longer veto windows if they can be afforded. 
In this analysis, the remaining SEs still dominate the electron spectra. 
Above the SE tail of 2.5~\el, the rate of electron events drops by a factor of $10^{4}$--$10^{5}$ and these values appear to be insensitive to changes of veto windows. 

Figure~\ref{fig:serate_vetoed} (right) shows the X-Y position distribution of multiple-electron (ME) pulses after the veto cuts. 
In contrast to the residual SEs that are approximately uniformly distributed in the X-Y space, MEs are observed to congregate at the edge and also some inner regions of the detector. 
A temporal study revealed that the majority of spatially congregated ME pulses in the inner volume emerged in bursts during short emission periods of 1--50 seconds. 
Similar transient electron emission from hot spots have also been reported in References~\cite{LZ2018_WireTest} and \cite{Bailey2016_Thesis}, and are attributed to grid emission. 
A total of three statistically significant rate spikes are identified in this 10-day data set, and additional ones with smaller amplitudes are observed but cannot be definitively differentiated from statistical fluctuations of the background rate. 
Given its frequent occurrence and varying amplitude, grid electron emission is expected to be responsible for a significant fraction of the residual electron events in LUX above the SE level. 

Although electron emission from metal surfaces is usually expected to produce SEs, electron multiplication can occur in high field regions~\cite{Alkhazov1970_EMultiplication} near physical of chemical defects, which explains the ME pulses in the intense grid emission periods. 
Such a multiplication effect was observed in LUX during a grid conditioning campaign that took place in 2014 between WS2013 and WS2014--16, when the high voltage on the LUX gate grid was increased to past the onset of intense electron emission~\cite{Bailey2016_Thesis}. 
During normal LUX operations, the grid voltage was reduced to avoid spurious electron emission and high voltage instabilities. 
However, this observation of electron rate spikes from hot spots suggests that the grid emission pathology still plays a significant role in LUX, and, if left unmitigated, possibly also in other experiments that appear to maintain otherwise stable high voltage operations. 

With an illustrative fiducial cut of $r<$12~cm, the ME rate in Figure~\ref{fig:serate_vetoed} (left) is reduced by another factor of 2, to 30--40 events/ton/day/\el. 
This rate is approximately 10 times higher than that reported by XENON1T at 4.5 extracted electrons~\cite{XENON1T2019_S2Only}, but we emphasize that this analysis, which focuses on the characterization of electron background in dual-phase xenon TPCs such as LUX, does not investigate all possible background rejection methods. 
For example, the shape of S2 pulses has been demonstrated to be a powerful tool in rejecting ME events from electrode grids~\cite{SORENSEN2011_Diffusion, XENON1T2019_S2Only}, so a dedicated ionization-only analysis with LUX data can lead to higher sensitivity to light dark matter interactions than that inferred from Figure~\ref{fig:serate_vetoed} (left). 
In addition, other mechanisms of ME emission from detector surfaces, such as radioactive decays from radon progenies (Sec.~\ref{sec:source}), are expected to exist, but the study of these background sources is beyond the scope of this work. 

\section{Summary}
\label{sec:concl}

From Sec.~\ref{sec:overview} to \ref{sec:me}, we described the major electron background pathologies observed in the LUX experiment, including both individual electron emission and multiple electron emission. 
By investigating the time, position and energy correlations between background electrons and energy depositions in the detector, 
we related these electrons to expected charge dynamics in LUX, and explained the likely emission mechanisms. 

Photoionization electron production by VUV light is a well-known phenomenon and has been discussed in liquid xenon TPCs~\cite{ZEPLIN2008_SE, XENON2014_SE}.  
The electron production rate by S1 and S2 light in LUX is measured to be (5--20)$\times10^{-5}$~\el/$\gamma$/m, which strongly depends on the concentration of electronegative impurities in liquid xenon. 
Instead of the previous hypothesis that negative ions such as O$_2^-$ may play a major role in the photoionization emission, our position and energy analysis suggests that some unknown neutral impurities are likely to dominate this process. 
Due to the exclusive SE production by photoionization and the close proximity of the SE detection time to the photon signal time, photoionization by large S1s or S2s is unlikely to be a significant background for ionization-based dark matter searches. 
On the other hand, their strong and robust correlation with electronegative impurities makes photoionization electrons a possible xenon purity monitor. 
We also evaluated the photoelectric efficiency of the LUX stainless steel grids at the xenon light wavelength, and obtained a result that is significantly higher than that measured for stainless steel in vacuum~\cite{Feuerbacher1971_MetalPE}, consistent with a reduction of its work function in liquid xenon~\cite{Tauchert1975_LXeGroundEnergy, Dowell2009_QE}. 

Due to the large electron-affinity of liquid xenon, complete extraction of ionization electrons in dual-phase xenon TPCs requires extremely high electric field~\cite{Xu2019_EEE}. 
In LUX, approximately 30--50\% of ionization electrons drifted to the liquid surface fail to be extracted and become temporarily trapped under the liquid surface. 
In this work, we observed strong evidence that these surface-trapped electrons are emitted in clusters at a timescale of milliseconds to tens of milliseconds. 
This process produces a distinct signature of  electron bursts that can contain a number of electrons in an extended period of up to a few milliseconds. 
We further propose that surface-trapped electrons that are not emitted in bursts are captured by electronegative impurities at the liquid surface, 
with an effective electron lifetime longer than that in the bulk liquid xenon. 
As a result, both the duration and the strength of e-burst emission exhibit an anti-correlation with the impurity level in liquid xenon.
Since multiple electrons may be simultaneously emitted from the same location, unextracted electrons can become a background for low-energy dark matter searches. 

Further, we report the observation and characterization of background electrons released by negatively charged impurities, 
which are formed by electronegative impurities capturing electrons in the liquid. 
This phenomenon produces a long-lasting SE background, the rate of which increases with the ionization energy deposited in the detector, the drift length of ionization electrons, and the impurity concentration in liquid xenon. 
This emission process approximately follows a power law-like time dependence, and results in electron emission up to seconds after the electron capture. 
Due to this large time delay, impurity-released electrons cannot be efficiently removed with a veto after high-energy events, and are found to dominate the residual electron spectrum in LUX in the single- to few-electron region. 

In addition to pathological electrons originating from preceding particle interactions with liquid xenon, we also observed background electrons that do not appear to correlate with energy depositions in the detector.
Mostly notably, pulses containing multiple electrons are detected from hot spots in the detector in short emission periods.
These electrons are attributed to grid emission in the TPC, which had been observed and studied in the LUX grid conditioning campaign when high voltage was raised to unstable levels~\cite{Bailey2016_Thesis}. 
This work demonstrates that cathodic grid emission can also occur during stable detector operation, 
and, if not properly mitigated~\cite{LZ2018_WireTest}, the produced multiple electrons pulses can be a significant background for ionization-only dark matter searches. 

This characterization of background electrons in LUX provides useful guidance for future experimental and analytical work that searches for low energy ionization-only interactions in dual-phase xenon TPCs. 
For an experiment that targets sensitivity in the single- to few-electron region, it is crucial to substantially suppress the number of electrons captured by impurities during drift. 
This may be achieved with a good liquid xenon purity, a short drift length, and a low ionization rate in the active region especially where the anticipated electron drifts are long. 
Possible experimental methods to improve the liquid xenon purity include eliminating high-outgassing materials from the TPC volume, isolating the clean active xenon from  TPC components that may outgas significantly~\cite{Sato2019_SealedTPC, LBECA2020}, and investigating novel purification techniques. 
It is worth noting that as the liquid xenon purity increases, delayed emission of surface-trapped electrons lasts longer, and a veto method as discussed in this analysis will become less efficient. 
Therefore, a high electron extraction efficiency close to 100\% will be beneficial~\cite{Xu2019_EEE}. 
To obtain a high extraction field and to mitigate grid emission in high electric field regions, surface treatments for the electrode grids, such as passivation, will be necessary~\cite{LZ2018_WireTest}. 
Residual grid emission as observed in this work can be identified by exploiting the temporal and spatial congregation of multi-electron pulses. 
For large detectors with a fine position resolution, veto cuts after high energy events can be implemented to part of the detector to avoid unnecessary loss of exposure, thanks to the strong position correlation between major background populations and preceding interactions. 
Future dual-phase xenon TPC experiments that implement these experimental and analytical methods can be expected to achieve high sensitivities in searches of low-mass dark matter interactions. 

\begin{acknowledgments}

This work was partially supported by the U.S. Department of Energy (DOE) under Award No. DE-AC02-05CH11231, DE-AC05-06OR23100, DE-AC52-07NA27344, DE-FG01-91ER40618, DE-FG02-08ER41549, DE-FG02-11ER41738, DE-FG02-91ER40674, DE-FG02-91ER40688, DE-FG02-95ER40917, DE-NA0000979, DE-SC0006605, DE-SC0010010, DE-SC0015535, and DE-SC0019066; the U.S. National Science Foundation under Grants No. PHY-0750671, PHY-0801536, PHY-1003660, PHY-1004661, PHY-1102470, PHY-1312561, PHY-1347449, PHY-1505868, and PHY-1636738; the Research Corporation Grant No. RA0350; the Center for Ultra-low Background Experiments in the Dakotas (CUBED); and the South Dakota School of Mines and Technology (SDSMT). 

Laborat\'{o}rio de Instrumenta\c{c}\~{a}o e F\'{i}sica Experimental de Part\'{i}culas (LIP)-Coimbra acknowledges funding from Funda\c{c}\~{a}o para a Ci\^{e}ncia e a Tecnologia (FCT) through the Project-Grant PTDC/FIS-NUC/1525/2014. Imperial College and Brown University thank the UK Royal Society for travel funds under the International Exchange Scheme (IE120804). The UK groups acknowledge institutional support from Imperial College London, University College London and Edinburgh University, and from the Science \& Technology Facilities Council for PhD studentships R504737 (EL), M126369B (NM), P006795 (AN), T93036D (RT) and N50449X (UU). This work was partially enabled by the University College London (UCL) Cosmoparticle Initiative. The University of Edinburgh is a charitable body, registered in Scotland, with Registration No. SC005336.\\

This research was conducted using computational resources and services at the Center for Computation and Visualization, Brown University, and also the Yale Science Research Software Core.\\

We gratefully acknowledge the logistical and technical support and the access to laboratory infrastructure provided to us by SURF and its personnel at Lead, South Dakota. SURF was developed by the South Dakota Science and Technology Authority, with an important philanthropic donation from T. Denny Sanford. 
SURF is a federally sponsored research facility under Award Number DE-SC0020216. 

J. Xu is partially supported by the U.S. DOE Office of Science, Office of High Energy Physics under Work Proposal Numbers SCW1676 and SCW1077 awarded to Lawrence Livermore National Laboratory (LLNL).
LLNL is operated by Lawrence Livermore National Security, LLC, for the U.S. Department of Energy, National Nuclear Security Administration under Contract DE-AC52-07NA27344.

\end{acknowledgments}

\bibliographystyle{apsrev}
\bibliography{biblio}

\end{document}